\documentstyle{amltd}
\input amssym.def
\input amssym.tex
\begin{document}
\annalsline{153}{2001}
\received{October 15, 1998}
\startingpage{149}
\def\bye{\end{document}}
 \font\tenrm=cmr10

\def\sni#1{\smallbreak\noindent{#1}. }
\def\ssni#1{\vglue-1pt\noindent\hskip18pt {#1}.}
\def\spn{\speqnu}
\def\nn{\nonumber}
\def\navs{\noalign{\vskip4pt}}
\def\ha{\noindent \phantom{mm}\hangindent=40pt\hangafter=1}
\catcode`\@=11
\font\twelvemsb=msbm10 scaled 1100
\font\tenmsb=msbm10
\font\ninemsb=msbm10 scaled 800
\newfam\msbfam
\textfont\msbfam=\twelvemsb  \scriptfont\msbfam=\ninemsb
  \scriptscriptfont\msbfam=\ninemsb
\def\msb@{\hexnumber@\msbfam}
\def\Bbb{\relax\ifmmode\let\next\Bbb@\else
 \def\next{\errmessage{Use \string\Bbb\space only in math
mode}}\fi\next}
\def\Bbb@#1{{\Bbb@@{#1}}}
\def\Bbb@@#1{\fam\msbfam#1}
\catcode`\@=12

 \catcode`\@=11
\font\twelveeuf=eufm10 scaled 1100
\font\teneuf=eufm10
\font\nineeuf=eufm7 scaled 1100
\newfam\euffam
\textfont\euffam=\twelveeuf  \scriptfont\euffam=\teneuf
  \scriptscriptfont\euffam=\nineeuf
\def\euf@{\hexnumber@\euffam}
\def\frak{\relax\ifmmode\let\next\frak@\else
 \def\next{\errmessage{Use \string\frak\space only in math
mode}}\fi\next}
\def\frak@#1{{\frak@@{#1}}}
\def\frak@@#1{\fam\euffam#1}
\catcode`\@=12

\title{Hermitian, symmetric and symplectic\\ random ensembles:
PDEs for the\\ distribution of the spectrum}

\shorttitle{Random ensembles} 

 \acknowledgements{The support of a National
Science Foundation grant \#DMS-98-4-50790 is
gratefully acknowledged.  
The support of a National Science
Foundation grant \#DMS-98-4-50790, a Nato, a FNRS and
a Francqui Foundation grant is gratefully
acknowledged.}
 \twoauthors{M. Adler}{P. van Moerbeke}
 \institutions{Brandeis University, Waltham, MA 02454
\\
{\eightpoint {\it E-mail address\/}:  adler@brandeis.edu}
\\
\vglue6pt
Universit\'e de Louvain, B1348 Louvain-la-Neuve, Belgium
\\
{\eightpoint {\it E-mail address\/}: vanm@geom.ucl.ac.be and
vanmoerbeke@brandeis.edu
}}

\catcode`\@=11
\let\c@equation=\relax
\newcounter{equation}[subsection]
\def\theequation{\thesubsection.\arabic{equation}}
\catcode`\@=12

\newcommand{\MAT}[1]{\left(\begin{array}{*#1c}}
\newcommand{\mat}{\end{array}\right)}
 
\newcommand{\rg}{\rightarrow}
\newcommand{\lrg}{\longrightarrow}
\newcommand{\Rg}{\Rightarrow}
\newcommand{\DF}{\Longleftrightarrow}
\newcommand{\pp}{\ldots}
\newcommand{\TT}{\tilde\tau}
\newcommand{\AR}{{\cal A}}
\newcommand{\CR}{{\cal C}}
\newcommand{\DR}{{\cal D}}
\newcommand{\FR}{{\cal F}}
\newcommand{\GR}{{\cal G}}
\newcommand{\HR}{{\cal H}}
\newcommand{\JR}{{\cal J}}
\newcommand{\LR}{{\cal L}}
\newcommand{\MR}{{\cal M}}
\newcommand{\NR}{{\cal N}}
\newcommand{\OR}{{\cal O}}
\newcommand{\PR}{{\cal P}}
\newcommand{\SR}{{\cal S}}
\newcommand{\VR}{{\cal V}}
\newcommand{\WR}{{\cal W}}
\newcommand{\BA}{{\Bbb A}}
\newcommand{\BC}{{\Bbb C}}
\newcommand{\BD}{{\Bbb D}}
\newcommand{\BF}{{\Bbb F}}
\newcommand{\BI}{{\Bbb I}}
\newcommand{\BX}{{\Bbb X}}
\newcommand{\BY}{{\Bbb Y}}
\newcommand{\BZ}{{\Bbb Z}}
\newcommand{\Sg}{\Sigma}
\newcommand{\iy}{\infty}
\newcommand{\pl}{\partial}
\newcommand{\al}{\alpha}
\newcommand{\vr}{\varepsilon}
 
\newcommand{\om}{\omega}
\newcommand{\vp}{\varphi}
\newcommand{\la}{\langle}
\newcommand{\ra}{\rangle}
\newcommand{\ga}{\gamma}
\newcommand{\Ga}{\Gamma}
\newcommand{\dt}{\delta}
\newcommand{\Dt}{\Delta}
\newcommand{\sg}{\sigma}
\newcommand{\BR}{{\Bbb R}}
\newcommand{\lb}{\lambda}
\newcommand{\Lb}{\Lambda}
\newcommand{\tr}{\mbox{tr}}

\newcommand{\BJ}{{\Bbb J}}
\newcommand{\Span}{\operatorname{span}}
\newcommand{\diag}{\operatorname{diag}}
\newcommand{\Res}{\operatorname{Res}}

\def\be#1\ee{\begin{equation}#1\end{equation}}
\def\bea#1\eea{\begin{eqnarray}#1\end{eqnarray}}
\def\bean#1\eean{\begin{eqnarray*}#1\end{eqnarray*}}

\newcommand{\Pf}{{\rm Pfaff}}
\newcommand{\Tr}{{\rm Tr}}
\newcommand{\Mat}{{\rm Mat}}
\newcommand{\OneOrTwo}{{\left\{\substack{1\\2}\right\}}}



\bigbreak
\centerline{\bf Abstract}\bigbreak
 Given the Hermitian, symmetric and
symplectic ensembles, it is shown that the probability
that the spectrum belongs to one or several intervals
satisfies a nonlinear PDE. This is done for the three
classical ensembles: Gaussian, Laguerre and Jacobi.
For the Hermitian ensemble, the PDE (in the boundary
points of the intervals) is related to the Toda
lattice and the KP equation, whereas for the symmetric
and symplectic ensembles the PDE is an inductive
equation, related to the so-called Pfaff-KP equation
and the Pfaff lattice. The method consists of
inserting time-variables in the integral and showing
that this integral satisfies integrable lattice
equations and Virasoro constraints.
 
\bigbreak

\centerline{\bf Contents}
\sni{0} Introduction
\ssni{0.1}  Hermitian, symmetric and symplectic Gaussian ensembles
\ssni{0.2}  Hermitian, symmetric and symplectic Laguerre  ensembles
\ssni{0.3}  Hermitian, symmetric and symplectic Jacobi ensembles
\ssni{0.4} ODEs, when $E$ has one boundary point

\sni{1} Beta-integrals
\ssni{1.1} Virasoro constraints for $\beta$-integrals
\ssni{1.2} Proof: $\beta$-integrals as fixed points of vertex operators
\ssni{1.3} Examples

\sni{2} Matrix integrals and associated integrable systems
\ssni{2.1} Hermitian matrix integrals and the Toda lattice
\ssni{2.2} Symmetric/symplectic matrix integrals and the Pfaff lattice

\sni{3} Expressing $t$-partials in terms of boundary-partials
\ssni{3.1} Gaussian and Laguerre ensembles
\ssni{3.2} Jacobi ensemble
\ssni{3.3} Evaluating the matrix integrals on the full range

\sni{4} Proof of Theorems 0.1, 0.2, 0.3
\ssni{4.1} $\beta=2,1$
\ssni{4.2} $\beta=4$, using duality
\ssni{4.3} Reduction to Chazy and Painlev\'e equations $(\beta=2)$

\sni{5} Appendix. Self-similarity proof of the Virasoro constraints  (Theorem 1.1)

\vglue24pt \centerline{\bf 0. Introduction}
\bigbreak

Consider weights of the form $\rho(z)dz:=e^{-V(z)}dz$
on an interval $F=[A,B]\subseteq\BR$, with rational
logarithmic derivative and subjected to the following
boundary conditions:
\be
-\frac{\rho'}{\rho}=V^{\prime}=\frac{g}{f}=
 \frac{\sum_0^{\iy}b_iz^i}{\sum_0^{\iy}a_iz^i},
\quad \lim_{z\rightarrow
A,B}f(z)\rho(z)z^k=0\mbox{\,\,for all\,\,}k\geq 0,\hskip12pt  \spn{0.0.1}
  \ee 
together with a disjoint union of intervals,
  \be
E=\bigcup_1^{r}~[c_{2i-1},c_{2i}]\subseteq F\subseteq
\BR. \spn{0.0.2}
 \ee
  The data (0.0.1) and (0.0.2) define an algebra of differential
operators
 \be{\cal
B}_k=\sum_1^{2r} c_i^{k+1}f(c_i)\frac{\pl}{\pl c_i} \spn{0.0.3}
 .\ee  
Let ${\cal H}_n,~{\cal S}_n~\mbox{and}
  ~{\cal T}_n$ denote the Hermitian
  ($M=\bar M^{\top}$), symmetric ($M=  M^{\top}$)
  and ``symplectic" ensembles ($M=\bar
                     M^{\top},~M=J\bar M J^{-1}
$), respectively. Traditionally, the latter is called
the ``symplectic ensemble," although the matrices
involved are not symplectic! These conditions
guarantee the reality of the spectrum of $M$. Then,
 ${\cal H}_n(E),~{\cal S}_n(E)~\mbox{and}
  ~{\cal T}_n(E)$ denote the subsets of
  ${\cal H}_n,~{\cal S}_n~\mbox{and}
  ~{\cal T}_n$ with
   spectrum in the subset $E\subseteq F\subseteq \BR$.
   The aim of this paper is to find PDEs for the
probabilities
\bea
&&\spn{0.0.4}\\
  P_n(E):&=& P_n(\mbox{ all spectral points of }~M
\in  E)\nonumber \\ \noalign{\vskip4pt}
  &=&\frac{
  \int_{{\cal H}_n(E),~{\cal S}_n(E)~\mbox{\tiny{or}}
  ~{\cal T}_n(E)}e^{-tr ~V(M)}dM}
  {\int_{{\cal H}_n (F),~{\cal S}_n(F)~\mbox{\tiny{or}}
  ~{\cal T}_n(F)}e^{-tr ~ V(M)}dM}
   \nonumber
  \\ \noalign{\vskip4pt} &=&
  \frac{\int_{E^n}|\Dt_n(z)|^{\beta}\prod_{k=1}^n
e^{-V(z_k)}dz_k}{\int_{F^n}|\Dt_n(z)|^{\beta}
 \prod_{k=1}^n e^{-V(z_k)}dz_k}, ~~~~~ \beta=2,1,4
~~\mbox{respectively}, \nonumber \eea
 for the Gaussian, Laguerre and
Jacobi weights. The probabilities involve parameters $\beta, a,
 b$ (see (0.1.1), (0.2.1) and (0.3.2)) and $$
\delta^{\beta}_{1,4}:=2
\left(\left(\frac{\beta}{2}\right)^{1/2}
-\left(\frac{\beta}{2}\right)^{-1/2}  \right)^2=
\left\{ \begin{array}{l} 0 ~~ \mbox{for}~~\beta=2\\  1
~~ \mbox{for}~~\beta=1,4.\\
\end{array}\right.
 $$

 The method used to obtain these PDEs involves
  inserting time-parameters into the integrals, appearing
  in (0.0.4) and to notice that the integrals obtained
  satisfy
\medbreak
  \noindent $\bullet$ Virasoro constraints: linear PDEs in $t$ and the boundary points of
 $E$, and 
  \medbreak
  \noindent $\bullet$ integrable hierarchies:
  $$\begin{array}{l|l|l} \mbox{ensemble} &~~~ \beta
   &\mbox{lattice}\\
  \hline
  \\
  \mbox{Hermitian} & \beta=2& \mbox{Toda}
  \\
  \mbox{symmetric} & \beta =1 & \mbox{Pfaff}\\
  \mbox{symplectic} & \beta =4& \mbox{Pfaff}
  \end{array} \quad \lower27pt\hbox{.} $$
As a consequence of a duality (explained in Theorem
1.1) between $\beta$-Virasoro generators under the map
$\beta \mapsto 4/\beta$, the PDEs obtained have a
remarkable property: the coefficients $Q$ and $Q_i$ in
the PDEs are functions of the variables
$n,\beta,a,b$, and have the invariance property under
the map
 $$
  n \rightarrow -2
n,~a \rightarrow -\frac{  a}{2},~b \rightarrow -\frac{
b}{2} ;
 $$
 to be precise,
  \be\left.
Q_i(-2n,\beta ,-\frac{a}{2},-\frac{b}{2}
)\right|_{\beta =1}=\left.Q_i(n, \beta,a,b
)\right|_{\beta =4}. \spn{0.0.5}
 \ee

\demo{Important remark} For $\beta=2$, the
probabilities satisfy PDEs
in the boundary points of $E$, whereas in the case
$\beta=1,4$, the equations are inductive. Namely, for
$\beta=1$ (resp.\ $\beta=4$), the probabilities
$P_{n+2}$ (resp.\ $P_{n+1}$) are given in terms of
$P_{n-2}$ (resp.\ $P_{n-1}$) and a differential
operator acting on $P_n$.
\enddemo
 
\demo{{\rm 0.1.}  Hermitian{\rm ,} symmetric and symplectic Gaussian
ensembles}
Given the disjoint union $E\subset \BR$ and the weight
$e^{-b z^2}$, the differential operators ${\cal
  B}_k$ take on the form
   $${\cal
  B}_k=\sum_1^{2r}c_i^{k+1}\frac{\pl}{\pl c_i}.$$
Also, define the {\em invariant} polynomials (in the
sense of  (0.0.5))
  \begin{eqnarray*} Q&=& 12
b^2n\left(n+1-\frac{2}{\beta}\right),\quad Q_2=4 (1+\delta^{\beta}_{1,4}) b
\left(2n+\delta^{\beta}_{1,4}
 (1-\frac{2}{\beta})\right)\\
\noalign{\noindent and} 
 Q_1&=& \left(2-\delta^\beta_{1,4} \right) {b^2\over \beta}.\end{eqnarray*}
\enddemo

\proclaim{Theorem}
The following probabilities for $(\beta=2,1,4)$
 \be P_n(E)= \frac{\int_{E^n}|\Dt_n(z)|^{\beta}\prod_{k=1}^n
e^{-b
z_k^2}dz_k}{\int_{\BR^n}|\Dt_n(z)|^{\beta}\prod_{k=1}^n
e^{-b z_k^2}dz_k}
,  \spn{0.1.1}\ee
  satisfy the {\rm PDE'}\/s $(F:=F_n=\log P_n)${\rm :}
\begin{eqnarray} &&
\spn{0.1.2}\\
&& \displaystyle{\delta^{\beta}_{1,4} Q
\left(\frac{P_{n-{2 \atop 1}}P_{n+{2 \atop 1}
}}{P_n^2}-1\right)}  \mbox{ with index }
\left\{
\begin{array}{l} 2 ~~ \mbox{when $n$ is even and}~~\beta=1\\
  1 ~~\mbox{when $n$ is arbitrary and}
~\beta=4\nn\\
\end{array}\right.   \\
&&\quad=\left({\cal B} _{-1}^4+(Q_2+6{\cal B}_{-1}^2F){\cal
B}_{-1}^2+4Q_1(3{\cal
B}^2_0-4
 {\cal B}_{-1}{\cal B}_1+6{\cal B}_0)\right)F.
\nn
\end{eqnarray}
\endproclaim
 
\demo{{\rm 0.2.} Hermitian{\rm ,} symmetric and symplectic Laguerre
ensembles} Given the disjoint union $E\subset \BR^+$
and the weight $z^{a}e^{-b z}$, the ${\cal B}_k$ take
on the form
  $${\cal
  B}_k=\sum_1^{2r}c_i^{k+2}\frac{\pl}{\pl c_i}.$$
Also define the polynomials, again respecting the
duality (0.0.5),
 \bean
  Q&=&\left\{\begin{array}{ll}
\displaystyle{\frac{3}{4} n(n-1)( n+2a)(
n+2a+1)},~~~\mbox{for}~~\beta=1\\  \\ \displaystyle{
\frac{3}{2}n(2n+1)(2n+a)(2n+a-1)}
,~~~\mbox{for}~~\beta=4
\end{array},\right.\\ \noalign{\vskip5pt}
Q_2&=&\left(
  3\beta n^{2}-\frac{a^2}{\beta}+6a n+4(1-\frac{\beta}{2})
  a
  +3\right)\delta^{\beta}_{1,4}+(1-a^2)(1-\delta^{\beta}_{1,4}),\\
\noalign{\vskip5pt}
  Q_1&=&\left(
  \beta n^{2}+2a n+(1-\frac{\beta}{2}) a\right),~~~~
  Q_0= b (2-\delta^{\beta}_{1,4})(n+
 \frac{a}{\beta}),
\\
Q_{-1}&=& {b^2\over \beta} \left(2-\delta^\beta_{1,4}\right).
  \eean
\enddemo

\proclaim{Theorem} 
   The following probabilities
 \be P_n(E)=\frac{\int_{E^n}|\Dt_n(z)|^{\beta}\prod_{k=1}^n
z_k^{a} e^{-b z_k}dz_k
}{\int_{\BR_+^n}|\Dt_n(z)|^{\beta}\prod_{k=1}^n
z_k^{a} e^{-b z_k}dz_k} \spn{0.2.1}
 \ee
  satisfy the {\rm PDE}\footnote{with the same convention on the
  indices
  $n\pm 2$ and $n\pm 1$, as in (0.1.2)}{\rm :  (}$F:=F_n=\log P_n${\rm )}
\begin{eqnarray}\noalign{\vskip-4pt}
&&\spn{0.2.2}\\
&&\hskip-24pt  \delta^{\beta}_{1,4} Q\left(\frac{P_{n-{2\atop 1}}P_{n+{2\atop 1}}}{P_n^2}-1 \right)\nonumber\\
 &&\enspace =\Bigl( {\cal B}_{-1}^4 -2(\delta^{\beta}_{1,4}
+1){\cal B}_{-1}^3 \nonumber\\
&&\qquad+ \ (Q_2+ 6 {\cal
B}_{-1}^2F-4(\delta^{\beta}_{1,4}+1) {\cal B}_{-1}F)
{\cal B}_{-1}^2 -3 \delta^{\beta}_{1,4} (Q_1-{\cal
B}_{-1}F)
  {\cal B}_{-1}\nonumber \\
&&\qquad  +\ Q_{-1} (3
 {\cal B}^2_{0}-4{\cal B}_{1}{\cal B}_{-1}
-2{\cal B}_{1})
  +Q_0(2{\cal B}_{0}{\cal B}_{-1}
    -{\cal B}_{0}) \Bigr)
   F. \nonumber 
 \end{eqnarray}
\endproclaim

\demo{{\rm 0.3.}  Hermitian{\rm ,} symmetric and symplectic Jacobi
ensembles}
  In terms of $E\subset [-1,1]$ and the Jacobi weight
$(1-z)^{a}(1+z)^{b}$, the differential
 operators ${\cal B}_k$ take on the form
   $${\cal
  B}_k=\sum_1^{2r}c_i^{k+1}(1-c_i^2)\frac{\pl}{\pl c_i}.$$
Setting $b_0=a-b,~b_1=a+b$, we introduce the new variables, which themselves have
 the invariance property (0.0.5):
 $$
  r=\frac{4}{\beta}(b_0^2+(b_1+2-\beta)^2)
%
~~~~~~s=\frac{4}{\beta}b_0(b_1+2-\beta)
$$ $$ q_n= \frac{4}{\beta}(\beta n+b_1 +2-\beta)(\beta n
 +b_1),
$$
 and the following  polynomials in $q=q_n,r,s$, thus {\it invariant} under the map (0.0.5):
 \begin{eqnarray}
\qquad Q&=&\frac{3}{16}\left(
(s^{2}-qr+q^{2})^{2}-4(rs^{2}-4q
 s^{2}-4s^{2}+q^{2}r)\right), \spn{0.3.1}
\\ \noalign{\vskip6pt}
  Q_1&=& 3s
 ^{2}-3qr-6r+2q^{2}+23q+24,\nn\\ \noalign{\vskip6pt}
 Q_2&=&3qs^{2}
 +9s^{2}-4q^{2}\,r+2qr+4q^{3}+10q^{2},
\nn \\ \noalign{\vskip6pt}
 Q_3&=&3qs^{2}+6s^{2}-3q^{2}
 r+q^{3}+4q^{2},
\nn \\ \noalign{\vskip6pt}
 Q_4&=& 9s
 ^{2}-3qr-6\,r+q^{2}+22q+24=Q_1+(6s^2-q^2-q).\nn 
 \end{eqnarray}
 \enddemo
 
\proclaim{Theorem}
   The following probabilities
 \be P_n(E)=\frac{\int_{E^n}|\Dt_n(z)|^{\beta}\prod_{k=1}^n
(1-z_k)^{a}(1+z_k)^{b}dz_k
}{\int_{[-1,1]^n}|\Dt_n(z)|^{\beta}\prod_{k=1}^n
(1-z_k)^{a}(1+z_k)^{b}dz_k}\hskip.75in \spn{0.3.2}
 \ee
  satisfy the {\rm PDE (}$F=F_n=\log P_n)$\/{\rm :}\/

  \bigbreak

 for $\beta=2${\rm :} 
 \bea 
&&\spn{0.3.3}\\
&&\Bigl(2 {\cal
B}_{-1}^4+(q-r+4){\cal B}^2_{-1}-(4{\cal B}_{-1}F-s)
{\cal B}_{-1}+3q {\cal B} _{0}^2 - 2q {\cal
B}_{0}+8{\cal B}_0 {\cal B}_{-1}^2 \nn\\ \noalign{\vskip6pt}
&&\quad-4(q-1){\cal B}_{1}{\cal B}_{-1} +(4{\cal
B}_{-1}F-s) {\cal B}_1 +2(4{\cal B}_{-1}F-s) {\cal
B}_{0} {\cal B}_{-1}+
 2 q {\cal B}_2 \Bigr)F
\nonumber\\ \noalign{\vskip6pt} && \hspace{2cm}
  +4{\cal B}^2_{-1}F\left( 2{\cal B}_{0}F
   +3{\cal B}^2_{-1}F\right) \nonumber
=0\eea

\pagebreak

    for $\beta=1,4${\rm :}  
\bigbreak

\noindent {\rm (0.3.4)} \quad {\ninepoint

 $\displaystyle{Q\left(\frac{P_{n+{2\atop
1}}P_{n-{2\atop 1}}} {P_n^2}-1\right)}$ \bea
\noalign{\vskip4pt}
&&=(q+1)\Bigl(4q {\cal
 B}^4_{-1}
 +12(4{\cal B}_{-1}F-s)
 {\cal B}^3_{-1}+2\left(q+12\right)(4{\cal B}_{-1}F-s)
 {\cal B}_{0}{\cal B}_{-1} \nonumber\\ \noalign{\vskip4pt}
 &&+3q^{2}
 {\cal B}_{0}^2 -4\left(q-4\right)q {\cal B}_{1}{\cal
B}_{-1}+q(4{\cal B}_{-1}F-s)  {\cal B}_{1}+20q{\cal
B}_{0}{\cal B}^2_{-1}+2q^{2}{\cal B}_{2}\Bigr) F
 \nonumber\\ \noalign{\vskip4pt}
 &&+\Bigl(Q_2{\cal B}^2_{-1}-sQ_1{\cal
 B}_{-1}+Q_3{{\cal B}_0}\Bigr)F +48({\cal B}_{-1}F)^{4}
  -48s({\cal
B}_{-1}F)^{3}+2Q_4({\cal B}_{-1}F)^{2} \nonumber\\ \noalign{\vskip4pt}
 &&+12\,q^{2}({{\cal
B}_0}F)^{2}+16\,q\,\left(2\,q-1\right)
 ({\cal B}^2_{-1}F)({{\cal B}_0}F)+24\left(q-1\right)q({\cal B}^2_{-1}F)^{2}
\nonumber \\ \noalign{\vskip4pt}
 &&+24  \Bigl(2{\cal B}_{-1}F
 -s\Bigr)\Bigl(  (q+2)
  {{\cal B}_0}F
 + (q+3){\cal B}^2_{-1}F\Bigr){\cal B}_{-1}F.\nn
\eea }

\endproclaim

 \demo{{\rm 0.4.} {\rm ODE}\/s{\rm ,} when $E$ has one boundary point}
Assume the set $E$ consists of one boundary point $c=x$,
besides the boundary of the full range. In that case
the PDEs in the previous section lead to ODEs in
$x$:

    (1) {\em Gaussian $(n\times n)$ matrix ensemble} 
 (for the function $\beta=2,1,4$):
 $$f_n(x)=\frac{d}{dx}\log P_n(\max_{i} ~
 \lb_i \leq x)$$
  satisfies
 \bea  
&& \delta^{\beta}_{1,4}Q\left(\frac{P_{n-{2\atop 1}}P_{n+{2\atop 1}}}{P_n^2}-1 \right)
\spn{0.4.1} \\
&&\qquad  = f_n^{\prime\prime\prime}+6 f_n^{\prime 2 }+
\left(4\frac{b^2x^2}{\beta}(\delta^{\beta}_{1,4}-2)
 +Q_2\right)f_n^{\prime
}-4\frac{b^2x}{\beta}(\delta^{\beta}_{1,4}-2)f_n.
\nonumber\eea

  (2) {\em Laguerre ensemble} (for $\beta=2,1,4$):
all eigenvalues $\lb_i$ satisfy $\lb_i\geq 0$ and
 $$f_n(x)=x\frac{d}{dx}\log P_n(\max_{i} ~
 \lb_i \leq x
 )$$
satisfies (with $f:=f_n(x)$)
 \bea
&&\spn{0.4.2}\\
 &&\hskip-24pt { \delta^{\beta}_{1,4}
Q\left(\frac{P_{n-{2\atop 1}}P_{n+{2\atop
1}}}{P_n^2}-1 \right)
 -\left(3\delta^{\beta}_{1,4} f
-\frac{b^2 x^2 }{\beta}(\delta^{\beta}_{1,4}-2)
-Q_0x-3 \delta^{\beta}_{1,4} Q_1\right)f }\nonumber\\
 &&\quad=
 x^3f^{\prime\prime\prime}-(2\delta^{\beta}_{1,4}
-1) x^2f^{\prime\prime}+6x^2f^{\prime 2} \nonumber\\
&&\qquad -x\left(4(\delta^{\beta}_{1,4}+1)f-
\frac{b^2x^2}{\beta}(\delta^{\beta}_{1,4}-2)
-2Q_0x-Q_2+2\delta^{\beta}_{1,4}+1\right) f^{\prime}.
\nonumber  \eea

 (3) {\em Jacobi ensemble}\/: all eigenvalues $\lambda_i$ satisfy
$-1 \leq \lb_i\leq 1$ and
$$f_n(x)=(1-x^2)\frac{d}{dx}\log P_n(\max_{i} \lb_i
\leq x)$$ satisfies (with $f:=f_n(x)$):

\bigbreak  
for {$\beta=2$}: \bea
 &&\hskip-24pt 2(x^2-1)^2f^{\prime\prime\prime}
 +4(x^2-1)\left(xf^{\prime\prime}
 -3f^{\prime 2}\right)
\spn{0.4.3}\\ \navs
&& +\left(16
xf-q(x^{2}-1)-2sx-r \right)f^{\prime} -f\left(4f-qx-s \right)=0\nonumber
 \eea

\bigbreak
for  {$\beta=1,4$}:
\medbreak

\noindent (0.4.4) \enspace {\ninepoint $\displaystyle{Q\left(\frac{P_{n+{2\atop
1}}P_{n-{2\atop 1}}} {P_n^2}-1\right)}$

\bea &=& 4(q+1)(x^2-1)^{2}\Bigl(-q(x^2-1)
f^{\prime\prime\prime}+
 (12f -qx-3s)
  f^{\prime\prime}+6 q(q-1)
 f^{\prime 2}\Bigr)\nonumber\\ \navs
  &&-\ (x^2-1)f^{\prime}\Bigl(24f(q+3)(2f-s)+8fq
  (5q-1)x-
   q(q+1)(qx^{2}+2sx+8)+
 Q_2
 \Bigr)
 \nonumber\\ \navs
 && +\ f\Bigl(48f^{3}+48f^{2}(qx+2x-s)+2f\left(8q^{2}x^{2}
 +2qx^{2}-12qsx-24s
 x+Q_4\right)\nonumber\\ \navs
  &&~~~~~~~~~-q(q+1)x(3qx^{2}+sx-2
 qx-3q)+Q_3x- Q_1s\Bigr).\nn \eea
 }

\enddemo 
For $\beta =2$, $f_n(x)$ satisfies a third-order
equation (of the so-called Chazy-type) with quadratic
nonlinearity in $f_n^{\prime}$. Then $f_n$ also
satisfies an equation, which is second-order in $f$
and quadratic in $f^{\prime\prime}$, which after some
rescaling can be put in a canonical form. Namely,
$$
\begin{array}{ll} 
{\rm Gauss} & g_n(z)=b^{-1/2}f_n(zb^{-1/2})+\frac{2}{3}nz,\\ \noalign{\vskip9pt}
{\rm Laguerre}, &
g_n(z)=f_n(z)+\frac{b}{4}(2n+a)z+\frac{a^2}{4},\\ \noalign{\vskip9pt}
{\rm Jacobi} &
g_n(z):=-\frac{1}{2}f_n(x)|_{x=2z-1}-\frac{q}{8}z+\frac{q+s}{16}
 \end{array}
 $$
satisfies the respective canonical equations of Cosgrove [11] and Cosgrove-Scoufis [12],
\bigbreak
\noindent $\bullet$ \enspace $  g^{\prime\prime 2}=-4g^{\prime 3
}+4(zg^{\prime}-g)^2+A_1g^{\prime}+A_2,$
\hfill ({Painlev\'e IV})
\bigbreak
\noindent$\bullet$ \enspace $ (zg^{\prime\prime })^2=(zg^{\prime}-g)\Bigl(-4g^{\prime 2
}+A_1(zg^{\prime}-g)+A_2\Bigr)+A_3g^{\prime}+A_4
, $
\hfill (Painlev\'e V)
\medbreak
\noindent $\bullet$ \enspace $ (z(z-1)g^{\prime\prime
})^2=(zg^{\prime}-g)\Bigl(4g^{\prime
2}-4g^{\prime}(zg^{\prime}-g)
 +A_2\Bigr)$ \hfill\smallbreak
\noindent\phantom{$\bullet$ \enspace $ (z(z-1)g^{\prime\prime
})^2=$} $+A_1g^{\prime
2}+A_3g^{\prime}+A_4, $
\hfill (Painlev\'e VI)
\medbreak\noindent 
with coefficients  which will be determined in Section
4.3. Each of these equations can be transformed into
the standard Painlev\'e equations.  

 For $\beta=1~\mbox{and}~4$,
 the inductive partial differential equations (0.1.2), (0.2.2)   and (0.3.4) are new. For $\beta =2$ and for general $E$, they
 were first computed by
 Adler-Shiota-van Moerbeke \cite{ASV1}, using the
method of the present paper. For $\beta=2$ and for $E$
having one boundary point, the equations obtained here
coincide with the ones first obtained by Tracy-Widom
in \cite{TW1}, who saw them to be Painlev\'e IV and V
for the Gaussian and Laguerre distribution
respectively. In his Louvain doctoral dissertation,
J.\ P. Semengue, together with L. Haine
 \cite{Haine}, were led to
 Painlev\'e VI for the Jacobi ensemble,
 for $\beta=2$ and $E$
 having one boundary point,
 upon subtracting the Tracy-Widom
 differential equation (\cite{TW1}) from the ones
 computed with the
 Adler-Shiota-van Moerbeke method (\cite{ASV1}).
  As we shall see, the classification of
   Cosgrove \cite{C} and Cosgrove-Scoufis \cite{CS},
(A.3) leads directly to these results.

 \section{Beta-integrals}

1.1. {\it Virasoro constraints for $\beta$\/{\rm -}\/integrals}.
Consider the data from (0.0.1) to (0.0.3) and the
 $t$-deformations of the
integrals (0.0.4), for {\em general} $\beta >0$:
($t:=(t_1,t_2,...)$ and $c=(c_1,c_2,...,c_{2r})$)
\be
I_n(t,c;\beta):=\int_{E^n}|\Dt_n(z)|^{\beta}\prod_{k=1}^n
\left(e^{\sum_1^{\iy}t_i z_k^i}\rho(z_k)dz_k\right)
~\mbox{for} ~~n>0.\hskip.35in\spn{1.1.1}\ee
 The main statement of this section is
 Theorem 1.1, whose proof will be outlined in the
 next subsection. In Section 5 (Appendix), we give a
 less conceptual proof, which is based on the invariance of the
 integral (1.1.2) below, under the transformation $z_i\mapsto
z_i+\vr f(z_i)z_i^{k+1}$ of the integration variables.
The central charge (1.1.6) has
 already appeared in the work of Awata et al.\
 \cite{Awata}.

\proclaimtitle{Adler-van Moerbeke \cite{AvM2}}
 \proclaim{Theorem} 
 The multiple integrals
 \be
I_n(t,c;\beta):=\int_{E^n}|\Dt_n(z)|^{\beta}\prod_{k=1}^n
\left(e^{\sum_1^{\iy}t_i z_k^i}\rho(z_k)dz_k\right)
~\mbox{for} ~~n>0 \hskip.35in \spn{1.1.2}
 \ee
 and
 \be
I_n(t,c;\frac{4}{\beta}):=\int_{E^n}|\Dt_n(z)|^{4/\beta}\prod_{k=1}^n
\left(e^{\sum_1^{\iy}t_i z_k^i} \rho(z_k)dz_k\right),
~\mbox{for} ~~n>0, \hskip.25in \spn{1.1.3}
 \ee
with $I_0=1${\rm ,} satisfy respectively the following
 Virasoro constraints\/\footnote{When $E$ equals the whole range $F$, then the
  ${\cal B}_k$'s are absent in the formulae (1.1.4).}
  for all $k\geq -1$\/{\rm :}\/
\bea
  &&   \left(-{\cal B}_k+ \sum_{i\geq
0}\left( a_i~ {}^{\beta}\BJ_{k+i,n}^{(2)}(t,n)-b_i ~
{}^{\beta} \BJ_{k+i+1,n}^{(1)}(t,n)\right)
\right)I_n(t,c;\beta) = 0,  \spn{1.1.4}  \\ && 
 \left(-{\cal B}_k + \sum_{i\geq 0}\left(
a_i~ {}^{\beta}\BJ_{k+i,n}^{(2)}\bigl(-\frac{\beta t
}{2},-\frac{2n}{\beta}\bigr) \right.\right.\nn\\
&&\hskip.5in\left.\phantom{\sum_{i+1} }\left.\left.+\frac{\beta b_i}{2} ~
{}^{\beta} \BJ_{k+i+1,n}^{(1)}\bigl(-\frac{\beta t
}{2},-\frac{2n}{\beta}\right)\right)
\right)I_n(t,c;\frac{4}{\beta}) = 0, \nonumber 
 \eea
  in terms of the coefficients
$a_i,~b_i$ of the rational function $(-\log \rho)'$
and the end points $c_i$ of the subset $E${\rm ,} as in
{\rm (0.0.1)} to {\rm (0.0.3).} For all $n\in \BZ${\rm ,} the
 $  {}^{\beta}\BJ_{k,n}^{(2)}(t,n) $
and $
 ~{}^{\beta}\BJ_{k,n}^{(1)}(t,n)$ form a Virasoro and a
Heisenberg algebra respectively{\rm ,} interacting as
follows\/{\rm :}\/
\bea \left[~ {}^{\beta}\BJ_{k,n}^{(2)},~
{}^{\beta}\BJ_{\ell,n}^{(2)} \right] &=&(k-\ell)~
{}^{\beta}\BJ_{k+\ell,n}^{(2)} +c\left(
\frac{k^3-k}{12} \right)\dt_{k,-\ell} \spn{1.1.5}\\ \noalign{\vskip4pt}
\left[~ {}^{\beta}\BJ_{k,n}^{(2)},~
{}^{\beta}\BJ_{\ell,n}^{(1)} \right] &=&-\ell ~ ~
{}^{\beta}\BJ_{k+\ell,n}^{(1)}+c'k(k+1)\delta_{k,-\ell} 
\nonumber\\ \noalign{\vskip4pt}  \left[~ {}^{\beta}\BJ_{k,n}^{(1)},~
{}^{\beta}\BJ_{\ell,n}^{(1)}  \right]
&=&\frac{k}{\beta}\delta_{k,-\ell},\nonumber \eea with central
charge
 \be
c=1-6\left(\left(\frac{\beta}{2}\right)^{1/2}
-\left(\frac{\beta}{2}\right)^{-1/2}  \right)^2 ~~~
\mbox{and}~~~c'=\left(\frac{1}{\beta}- \frac{1}{2}
\right).\hskip.5in \spn{1.1.6}\ee
\endproclaim

\demo{{R}emark {\rm 1}}
The $ {}^{\beta}\BJ_{k,n}^{(2)}$'s are defined as
follows:
 \be
 {}^{\beta}\BJ_{k,n}^{(2)}=\frac{\beta}{2}\sum_{i+j=k}
:~ {}^{\beta}\BJ_{i,n}^{(1)} ~
{}^{\beta}\BJ_{j,n}^{(1)}:
+\left(1-\frac{\beta}{2}\right)\left((k+1)~
{}^{\beta}\BJ_{k,n}^{(1)} -k\BJ_{k,n}^{(0)}\right)
.\spn{1.1.7}\ee
 Componentwise, we have $$ ~
{}^{\beta}\BJ_{k,n}^{(1)}(t,n)= ~
{}^{\beta}J_k^{(1)}+nJ_k^{(0)}
 ~\mbox{and}~ ~ {}^{\beta}\BJ_{k,n}^{(0)}=nJ_k^{(0)}= n\dt_{0k}
$$
and hence
\bean
  ~{}^{\beta}\BJ_{k,n}^{(2)}(t,n)
 &=&\Bigl(\frac{\beta}{2}\Bigr) ~~
{}^{\beta}J_k^{(2)} + \Bigl(n\beta
+(k+1)(1-\frac{\beta}{2})\Bigr) ~~ {}^{\beta}J_k^{(1)}
\\
&&+\ n\Bigl((n-1)\frac{\beta}{2}+1\Bigr)
J_k^{(0)},
 \eean
  where
  \bea
 ~ {}^{\beta}J_k^{(1)}&=&\frac{\pl}{\pl
t_k}+\frac{1}{\beta}(-k)t_{-k}
\spn{1.1.8}
 \\ \navs
  ^{\beta}J^{(2)}_{k}&=&\sum_{i+j=k}\frac{\pl^2}{\pl
 t_{i}\pl t_{j}}+\frac{2}{\beta}\sum_{-i+j=k}it_{i}\frac{\pl}{\pl
 t_{j}}+\frac{1}{\beta^2}\sum_{-i-j=k}it_{i}jt_{j}.
\nonumber \eea
We put $n$ explicitly in
 $ {}^{\beta}\BJ_{\ell,n}^{(2)}(t,n)$ to indicate that
  the
 $n^{\rm th}$ component contains $n$ explicitly, besides $t$.
\enddemo
\vglue6pt
\demo{{R}emark {\rm 2}}
The Heisenberg and Virasoro generators
satisfy the following {\em duality} properties:
  \bea
 {}^{\frac{4}
{\beta}  }\BJ_{\ell,n}^{(2)}\Bigl(t,n \Bigr)
&=&{}^{\beta}\BJ_{\ell,n}^{(2)}\left(-\frac{\beta
t}{2},-\frac{2n}{\beta}\right ) ,~n\in \BZ\spn{1.1.9}\\
 {}^{\frac{4} {\beta} }\BJ_{\ell,n}^{(1)}\Bigl(t,n
\Bigr) &=&-\frac{\beta}{2}~~
{}^{\beta}\BJ_{\ell,n}^{(1)}\left(-\frac{\beta
t}{2},-\frac{2n}{\beta}\right ),~n>0.\nn
  \eea
 In (1.1.9),
${}^{\beta}\BJ_{\ell,n}^{(2)}\left(- \beta
t/2,-2n/\beta \right )$ means that the variable $n$,
which appears in the $n^{\rm th}$ component, gets replaced by
$-2n/\beta$ and $t$ by $-\beta t/2$.
\enddemo

\demo{{\rm 1.2.} Proof\/{\rm :} $\beta$\/{\rm -}\/integrals as fixed points of vertex
operators}
The most transparent way to prove Theorem 1.1 is via
vector vertex operators, for which the
$\beta$-integrals are fixed points. This is a
technique which has been used by us  already in
\cite{AvM1}. Indeed, define the (vector) vertex
operator $\Bbb X$, for $t=(t_1,t_2,...)\in \BC^{\iy},~u \in
\BC$:
\be
\BX_{\beta}(t,u)=\Lb^{-1}e^{\sum_1^{\iy}t_i u^i}
e^{-\beta\sum_1^{\iy}\frac{ u^{-i}}{i} \frac{\pl}{\pl
t_i}}\chi(|u|^{\beta}), \spn{1.2.1}\ee 
where $\chi(z):= (1,z,z^2,\ldots)$.  The vertex operator   acts on vectors
$f(t)=\break (f_0(t),f_1(t),...)$ of functions,
 as follows\footnote{For $\alpha \in {\scriptstyle\BC}$, define $[\alpha]:=(\alpha, \frac{\alpha^2}{2},
 \frac{\alpha^3}{3},...) \in {\scriptstyle \BC}^{\iy}$. The operator $\Lb$ is the
 shift matrix, with zeroes everywhere, except for $1$'s just above
 the diagonal, i.e., $(\Lambda v)_n=v_{n+1}$. }
$$
\bigl(\BX_{\beta}(t,u)f(t)\bigr)_n=e^{\sum_1^{\iy}t_i
u^i}\left(|u|^{\beta }\right)^{n-1} f_{n-1}(t-\beta
[u^{-1}]). $$ For the sake of convenience, in  this section we  introduce the following vector Virasoro
generators: $~ {}^{\beta}\BJ_k^{(i)}(t):=~ ( ~
{}^{\beta}\BJ_{k,n}^{(i)}(t,n))_{n\in \BZ} $.
\enddemo

\vglue6pt

\proclaim{Proposition} The multiplication operator $z^k$ and the
differential operators $\frac{\pl}{\pl z}z^{k+1}$ with
$z \in \BC^*${\rm ,} acting on the vertex operator
$\BX_{\beta}(t,z)${\rm ,} have realizations as commutators{\rm ,}
in terms of the Heisenberg and Virasoro generators $~
{}^{\beta}\BJ_k^{(1)}(t)$ and $~
{}^{\beta}\BJ_k^{(2)}(t)${\rm :} \bea
z^{k}\BX_{\beta}(t,z)&=& \left[ ~
{}^{\beta}\BJ_k^{(1)}(t), \BX_{\beta}(t,z)\right],
\spn{1.2.2}\\ \navs
\noalign{\vskip4pt} \frac{\pl}{\pl
z}z^{k+1}\BX_{\beta}(t,z)&=& \left[~ {}^{\beta}
\BJ_k^{(2)}(t), \BX_{\beta}(t,z)\right]. \nn\eea
\endproclaim
 
\vglue6pt

\proclaim{{C}orollary}
 Given a weight $\rho(z)dz$ on $\BR$ satisfying {\rm (0.0.1),} we
 have
 \be
 \frac{\pl}{\pl z}z^{k+1}f(z)\BX_{\beta}(t,z)\rho(z)
 =
\left[ \sum_{i\geq 0}\left( a_i ~
{}^{\beta}\BJ_{k+i}^{(2)}(t)-b_i
 ~ {}^{\beta}\BJ_{k+i+1}^{(1)}(t)\right)
 , \BX_{\beta}(t,z)\rho(z)\right].\spn{1.2.3}
\ee
\endproclaim

\demo{Proof} Using (1.2.2) in the last line, compute
\begin{eqnarray}
  &&\quad \frac{\pl}{\pl z}
z^{k+1}f(z)\BX_{\beta}(t,z)\rho(z)\spn{1.2.4}\\ \noalign{\vskip6pt}
  &&\qquad =\ \left(\frac{\rho'(z)}{\rho(z)}f(z)
\right)z^{k+1}\BX_{\beta}(t,z)\rho(z)+
 \rho(z)\frac{\pl}{\pl z}\left(
z^{k+1}f(z)\BX_{\beta}(t,z)\right) \nonumber\\ \noalign{\vskip4pt}
&&\qquad =\ -\left(\sum_0^{\iy} b_i z^{k+i+1} \BX_{\beta}(t,z)
\right)\rho(z)+\rho(z)\frac{\pl}{\pl z}
 \left(\sum_0^{\iy} a_i z^{k+i+1} \BX_{\beta}(t,z)
\right) \nonumber\\ \noalign{\vskip4pt} &&\qquad =\ -\left[\sum_0^{\iy} b_i
~{}^{\beta}\BJ_{k+i+1}^{(1)},
  \BX_{\beta}(t,z)\rho(z) \right]
 +\left[\sum_0^{\iy} a_i ~{}^{\beta}\BJ_{k+i}^{(2)}
,\BX_{\beta}(t,z)\rho(z)\right], \nonumber
\end{eqnarray}
establishing (1.2.3).\enddemo

\vglue6pt

Given the weight $\rho_E(u)du=\rho(u)I_E(u)du$, with
$\rho$ and $E$ as before, and with $I_E$
 the indicator function of $E$, define the integrated vector
vertex operator
\be
\BY_{\beta}(t, \rho_E):= \int_{E} du
\rho(u)\BX_{\beta}(t,u),\spn{1.2.5} \ee and the vector operator
\bea \DR_k&:=&{\cal B}_k~~~-~~~\VR_k\spn{1.2.6}\\
&:=&\sum_1^{2r} c_i^{k+1}f(c_i)\frac{\pl}{\pl c_i}-
\sum_{i\geq 0}\left( a_i ~{}^{\beta}
\BJ_{k+i}^{(2)}(t)-b_i
 ~{}^{\beta}\BJ_{k+i+1}^{(1)}(t)\right),\nonumber
\eea consisting of
 a $c$-dependent boundary part ${\cal B}_k$ and a $(t,n)$-dependent Virasoro
 part $\VR_k$.

\proclaim{Proposition} The following commutation relation holds\/{\rm :}\/
\be
\left[\DR_k,\BY_{\beta}(t, \rho_E)\right]=0. \spn{1.2.7}\ee
\endproclaim

\demo{Proof} Integrating both sides of (1.2.3) over $E$,
one computes:
\begin{eqnarray}
&&\spn{1.2.8}\\
\int_Edz\frac{\pl}{\pl z}\left( z^{k+1}f(z)\BX_{\beta}
(t,z) \rho(z)\right) &=& \sum_1^{2r}
(-1)^ic_i^{k+1}f(c_i)
 \BX_{\beta}(t,c_i) \rho(c_i)\nn\\ \navs
 &=& \sum_1^{2r}  c_i^{k+1}f(c_i)
 \frac{\pl}{\pl c_i}\int_{E}
 \BX_{\beta}(t,z) \rho(z)dz\nonumber\\ \navs
 &=& \left[ {\cal B}_k, \BY_{\beta}(t,\rho_E) \right];\nonumber
\end{eqnarray}
while on the other hand
\begin{eqnarray}
&& \hskip-12pt\int_E dz \left[\sum_{i\geq
0}\left( a_i ~{}^{\beta}\BJ_{k+i}^{(2)}-b_i
 ~{}^{\beta}\BJ_{k+i+1}^{(1)}\right),\BX_{\beta}
(t,z) \rho(z)\right]\spn{1.2.9}\\ \noalign{\vskip4pt}
 &&\quad=\ \left[\sum_{i\geq 0}\left( a_i ~{}^{\beta}\BJ_{k+i}^{(2)}-b_i
 ~{}^{\beta}\BJ_{k+i+1}^{(1)}\right),\int_{E} dz\rho (z) \BX_{\beta}
(t,z) \right]\nonumber\\ \noalign{\vskip4pt}
 &&\quad =\ \left[ \VR_k, \BY_{\beta}(t, \rho_E) \right].\nn
\end{eqnarray}
 Subtracting both expressions (1.2.8) and
(1.2.9) yields, using (1.2.3),
  $$ 0=\left[{\cal B}_k- \VR_k,
\BY_{\beta}(t, \rho_E) \right]
 = \left[\DR_k, \BY_{\beta}(t, \rho_E) \right],
 $$
concluding the proof of Proposition 1.4.\enddemo 

\vglue8pt

\proclaim{Proposition}
The column vector{\rm ,}
$$I(t):=\left(\int_{E^n}|\Dt_n(z)|^{\beta}\prod_{k=1}^n
e^{\sum_1^{\iy}t_i z_k^i}\rho(z_k)dz_k\right)_{n\geq
0}$$ is a fixed point for the vertex operator
$\BY_{\beta}(t, \rho_E)${\rm :}
\be
\left(\BY_{\beta}(t, \rho_E)I\right)_n=I_n,~ n\geq 1.
\spn{1.2.10}\ee
\endproclaim

\demo{Proof} We have
 \bea
&&\spn{1.2.11}\\
I_n(t)&=&\int_{\BR^n}|\Dt_n(z)|^{\beta}
 \prod_{k=1}^n
\left(e^{\sum_1^{\iy}t_i
z_k^i} \rho_E(z_k)dz_k\right) \nn \\
&&\nonumber\\
 &=&\int_{\BR}du\rho_E(u)e^{\sum_1^{\iy}t_i u^i}
 |u|^{\beta (n-1)}\nonumber\\
 &&~~~~~~~\int_{\BR^{n-1}}
 \prod_{k=1}^{n-1}\left|1-\frac{z_k}{u}\right|^{\beta}
  |\Dt_{n-1}(z)|^{\beta}\prod_{k=1}^{n-1}
\left(e^{\sum_1^{\iy}t_i z_k^i}
\rho_E(z_k)dz_k\right)\nonumber\\
&=&\int_{\BR}du\rho_E(u)e^{\sum_1^{\iy}t_i u^i}
 |u|^{\beta (n-1)}\nonumber\\
&&~~~~e^{-\beta
\sum_1^{\iy}\frac{u^{-i}}{i}\frac{\pl}{\pl t_i}}
\int_{\BR^{n-1}}|\Dt_{n-1}(z)|^{\beta}
\prod_{k=1}^{n-1} \left(e^{\sum_1^{\iy}t_i z_k^i}
\rho_E(z_k)dz_k\right)\nonumber\\
&=&\int_{\BR}du\rho_E(u)|u|^{\beta
(n-1)}e^{\sum_1^{\iy}t_i u^i}
  e^{-\beta \sum_1^{\iy}\frac{u^{-i}}{i}
\frac{\pl}{\pl t_i}}I_{n-1}(t)\nonumber\\
&&\nonumber\\
 &=&\Big(\BY_{\beta}(t, \rho_E)I(t)\Big)_n.\nn
\eea
 It suffices to do the above argument for all $t_i>0$,
 enabling one to replace $e^{\sum_1^{\iy}t_i
z^i}$ by $|e^{\sum_1^{\iy}t_i
z^i}|$. Then one continues the result
for all $t_i\in \BC$.  \enddemo

\demo{Proof of Theorem {\rm 1.1}} From
Proposition 1.4 it follows that for $n\geq 1$,
  \bea
0&=&\left[\DR_k,\left( \BY_{\beta}(t,\rho_E) \right)^n
\right]I \spn{1.2.12}\\
\noalign{\vskip4pt} &=&\DR_k \BY_{\beta}(t,\rho_E)^n
I- \BY_{\beta}(t,\rho_E)^n\DR_k I.
\nn\eea Taking the $n^{\rm th}$ component for $n\geq 1$ and
$k\geq -1$, setting
 $$X_{\beta}(t,u)=e^{\sum t_i u^i}
 e^{-\beta \sum \frac{u^{-i}}{i}\frac{\pl}{\pl t_i}},
$$
 and using (1.2.10), we have
\bea 0&=& \left(\DR_k I-\BY_{\beta}(t,\rho_E)^n\DR_k
 I \right)_n\nonumber\\
 &=&(\DR_k I )_n- \int du\rho_E(u)X_{\beta}(t;u)
 (|u|^{\beta})^{n-1}...\int du\rho_E(u)X_{\beta}(t;u)
 (\DR_k I)_0 \nonumber\\
 &=& (\DR_k I)_n. \nonumber
\eea Indeed $(\DR_k I)_0=0$ for $k\geq -1$, since
$I_0=1$ and $\DR_k$ involves ${\cal B}_k,
{}^{\beta}J_{k}^{(2)},
 ~ {}^{\beta}J_{k}^{(1)}$ and $J_{k}^{(0)}$ for
  $k\geq -1$:
\medbreak
$\left\{\begin{array}{l}
 {\cal B}_k  \hbox{ and } {}^{\beta}J_{k}^{(2)}\mbox{ are pure
differentiations for $k\geq -1$;}\hspace{7.9cm} \\ \navs
 {}^{\beta} J_{k}^{(1)}\mbox{is pure
differentiation,
 except for $k=-1$; 
}
 \\
\navs \mbox{${}^{\beta}J_{-1}^{(1)}$ appears with
coefficient $n\beta$, which vanishes for
 $n=0$;}
  \\ \navs
 \mbox{$J_{k}^{(0)}$ appears with coefficient}~
 n((n-1)\frac{\beta}{2} +1),~\mbox{vanishing for
 $n=0$.}
\end{array}    \right.
   $
\medbreak
The proof of the $2^{\rm nd}$ formula in (1.1.4) follows
immediately from the duality (1.1.9).\enddemo

\demo{{\rm 1.3.} Examples.
 Example {\rm 1  (}Gaussian $\beta$\/{\rm -}\/integrals{\rm )}}
 The weight and the $a_i$ and $b_i$, as in
(0.0.1), are given by (setting $b=1$ in (0.1.1))
 $$\rho(z)=e^{-V(z)}= e^{-z^2},
~~V'=g/f=2z,$$ $$ a_0=1,b_0=0,b_1=2,~\mbox{and all
other}~a_i,b_i=0 .$$
  From Theorem 1.1, the integrals
\be
I_n= \int_{E^n}|\Dt_n(z)|^{\beta}\prod_{k=1}^n
e^{-z_k^2+\sum_{i=1}^{\iy}t_iz_k^i}dz_k \spn{1.3.1}
 \ee
satisfy the Virasoro constraints
 \be
  -{\cal B}_{k}I_n=
-\sum_1^{2r}c_i^{k+1}\frac{\pl}{\pl c_i}I_n=
 \left(-~{}^{\beta}\BJ_{k,n}^{(2)}+2
~{}^{\beta}\BJ_{k+2,n}^{(1)}\right)I_n,~~k=-1,0,1,\ldots \, . \spn{1.3.2}
  \ee
 Introducing the following notation
 $$ \sigma_i=
  (n-\frac{i+1}{2})\beta+i+1-b_0=
  (n-\frac{i+1}{2})\beta+i+1,
 $$
and upon setting $F=\log I_n$ we find that the first three constraints have the following
form:
 \bean
 -{\cal B}_{-1} F&=&
 \left(
2\frac{\pl}{\pl t_1}-\sum_{i\geq 2} it_i
\frac{\pl}{\pl t_{i-1}}  \right)F - nt_1 ,\\ \noalign{\vskip6pt}
 - {\cal B}_0 F  &=&
 \left(
2\frac{\pl}{\pl t_2}- \sum_{i\geq 1} it_i
\frac{\pl}{\pl t_i} \right)F - \frac{
n}{2}\sigma_1,
\\ \noalign{\vskip6pt}
 -{\cal B}_1 F &=&
 \left(
2\frac{\pl}{\pl t_3} -\sigma_1
\frac{\pl}{\pl t_1}- \sum_{i\geq 1} it_i
\frac{\pl}{\pl t_{i+1}} \right)F. \eean
\medbreak
 For later use, take linear combinations
 such that each expression contains the pure
 differentiation term $\pl F/\pl t_i$:
\be
{\cal D}_1=-\frac{1}{2}{\cal B}_{-1},~~{\cal
D}_2=-\frac{1}{2}{\cal B}_0,~~{\cal
D}_3=-\frac{1}{2}\left({\cal B}_1+\frac{\sigma_1
}{2}{\cal B}_{-1}\right), \hskip.5in\spn{1.3.3} \ee
which  yields
  \bea
 \qquad  {\cal
D}_1 F&=&\left(\frac{\pl}{\pl t_1}-\frac{1}{2}
\sum_{i\geq 2}it_i\frac{\pl}{\pl
t_{i-1}}\right)F-\frac{nt_1}{2},\spn{1.3.4} \\ \noalign{\vskip6pt}
{\cal D}_2
F&=&\left(\frac{\pl}{\pl t_2}-\frac{1}{2} \sum_{i\geq
1}it_i\frac{\pl}{\pl
t_i}\right)F-\frac{n}{4}\sigma_1,
\nonumber\\
\noalign{\vskip6pt} {\cal D}_3 F&=&\left(\frac{\pl}{\pl
t_3}-\frac{1}{2} \sum_{i\geq 1}it_i\frac{\pl}{\pl
t_{i+1}}-\frac{1}{4}\sigma_1 
\sum_{i\geq 2}it_i\frac{\pl}{\pl t_{i-1}}\right)F
-\frac{n}{4}\sigma_1 t_1 
 .\nonumber 
\eea

\enddemo

\demo{Example {\rm 2   (}Laguerre $\beta$-integrals{\rm )}}
 Here, the
weight and the $a_i$ and $b_i$, as in (0.0.1), are
given by (again setting $b=1$ in (0.2.1))
 $$e^{-V}=z^ae^{-z},~~V'=\frac{g}{f}=\frac{z-a}{z}, $$
 $$ a_0=0,~a_1=1,~ b_0=-a,~b_1=1,~\mbox{and all
other}~a_i,b_i=0 .$$ Thus from (1.1.4), the integrals
 \be
I_n= \int_{E^n}|\Dt_n(z)|^{\beta}\prod_{k=1}^n
z_k^{a}e^{-z_k+\sum_{i=1}^{\iy}t_iz_k^i}dz_k
 \spn{1.3.5} \ee
satisfy the Virasoro constraints, for $k\geq -1$,
 \be
- {\cal B}_k I_n=- \sum_1^{2r}c_i^{k+2}\frac{\pl}{\pl
c_i}I_n=\left(-~{}^{\beta}\BJ_{k+1,n}^{(2)} -a
~{}^{\beta}\BJ_{k+1,n}^{(1)}+~{}^{\beta}\BJ_{k+2,n}^{(1)}
 \right) I_n.\quad\spn{1.3.6} \ee
Introducing the following notation, as before,
 \bean \sigma_i&=& 
  (n-\frac{i+1}{2})\beta+i+1-b_0=(n-\frac{i+1}{2})\beta+i+1+a, 
\eean
and upon setting $F=F_n=\log
I_n$, we see that  the first three have the form:
 \begin{eqnarray*}
- {\cal B}_{-1} F&=& \left(\frac{\pl}{\pl
t_1}-\sum_{i\geq 1} it_i \frac{\pl}{\pl t_{i}}
 \right) F - \frac{n}{2}(\sigma_1+a),\\ \navs
- {\cal B}_0 F   &=&   \left(\frac{\pl}{\pl t_2} -
\sigma_1\frac{\pl}{\pl t_1}-\sum_{i\geq 1} it_i
 \frac{\pl}{\pl t_{i+1}} \right) F, \\ \navs
  -{\cal B}_1 F  &=& \left(\frac{\pl}{\pl
t_3}-\sigma_2\frac{\pl}{\pl t_2}-\sum_{i\geq 1} it_i
\frac{\pl}{\pl t_{i+2}}- \frac{\beta
}{2}\frac{\pl^2}{\pl t_1^2} \right) F 
 -
\frac{\beta }{2}\left(\frac{\pl F}{\pl t_1} \right)^2.
 \end{eqnarray*}

Replacing the operators ${\cal B}_i$ by linear
combinations ${\cal D}_i$, we see that
 \bea
 {\cal D}_1&=&-{\cal
B}_{-1}\spn{1.3.7}\\
 {\cal D}_2&=&-{\cal B}_0-\sigma_1{\cal B}_{-1}\nonumber\\
 {\cal D}_3&=&-{\cal B}_1-
 \sigma_2{\cal B}_0 -\sigma_1\sigma_2{\cal B}_{-1}\nonumber 
  \eea
  yields expressions, each containing a pure derivative
$\pl F/\pl t_i$
 \bea
&&\spn{1.3.8}\\
{\cal D}_1 F&=&\frac{\pl F}{\pl t_1}-\sum_{i\geq 1}
it_i\frac{\pl F}{\pl
t_i}-\frac{n}{2}(\sigma_1+a),\nn\\
 {\cal D}_2F&=&\frac{\pl F}{\pl
t_2}+ \sum_{i\geq 1} it_i \left(
-\sigma_1\frac{\pl}{\pl t_i}-\frac{\pl}{\pl
t_{i+1}}\right)F-\frac{n}{2}(\sigma_1+a)\sigma_1,
\nonumber\\
 {\cal D}_3F&=&\frac{\pl F}{\pl t_3} -\sum_{i\geq 1}it_i
\Biggl(\sigma_1\sigma_2\frac{\pl}{\pl t_i}
+\sigma_2\frac{\pl}{\pl t_{i+1}}+\frac{\pl}{\pl
t_{i+2}}\Biggr)F
-\frac{n}{2}(\sigma_1+a)\sigma_1\sigma_2 \nonumber\\   &&
 -\frac{\beta}{2}\left(\frac{\pl^2F}{\pl t^2_1}+
\left(\frac{\pl F}{\pl t_1}\right)^2\right).\nn
 \eea
\enddemo
 
\demo{Example {\rm 3  (}Jacobi $\beta$-integral\/{\rm )}}
 The
weight and the $a_i$ and $b_i$, as in (0.0.1), are
given by
 $$
\rho_{ab}(z):=e^{-V}=(1-z)^{a}(1+z)^{b} ,
V'=\frac{g}{f}=\frac{a-b+(a+b)z}{1-z^2},
 $$
 $$
 a_0=1,a_1=0,a_2=-1,b_0=a-b,b_1=a+b
 ,~\mbox{and all
 other}~a_i,b_i=0 .$$
 The integrals
 \be
 \int_{E^n}|\Dt_n(z)|^{\beta}\prod_{k=1}^n
(1-z_k)^{a}(1+z_k)^{b}e^{\sum_{i=1}^{\iy}t_iz_k^i}dz_k \spn{1.3.9}
 \ee
satisfy the Virasoro constraints $(k\geq-1)$:
 \bea 
   \qquad- {\cal B}_k I_n&=& -
\sum_1^{2r}c_i^{k+1}(1-c_i^2)\frac{\pl}{\pl c_i}I_n
\spn{1.3.10}\\   &=&
\left(~{}^{\beta}\BJ_{k+2,n}^{(2)}-~{}^{\beta}\BJ_{k,n}^{(2)}+
b_0~{}^{\beta}\BJ_{k+1,n}^{(1)}
+b_1~{}^{\beta}\BJ_{k+2,n}^{(1)}\right)I_n. \nn\eea
 Introducing the following notation, 
  \bean \sigma_i&=&(n-\frac{i+1}{2})\beta+i+1+b_1, 
\eean
and upon setting $F=F_n=\log
I_n$, we see that  the first four have the following form:
 \bea
&&\spn{1.3.11}\\
 - {\cal B}_{-1} F&=& \left(\sigma_1
 \frac{\pl}{\pl
t_1}+\sum_{i\geq 1} it_i \frac{\pl}{\pl t_{i+1}}-
 \sum_{i\geq 2} it_i \frac{\pl}{\pl t_{i-1}}
 \right) F 
 +n(b_0-t_1) ,\nn \\  &&  \nonumber\\
 -{\cal B}_0 F   &=&   \left(
 \sigma_2 \frac{\pl}{\pl t_2}
+  b_0\frac{\pl}{\pl t_1}
 +\sum_{i\geq 1}it_i
 (\frac{\pl}{\pl t_{i+2}}-\frac{\pl}{\pl t_{i}})
   +\frac{\beta}{2}\frac{\pl^2 }{\pl t_1^2}
   \right) F \nonumber\\  
 &&+\frac{\beta}{2}\left(\frac{\pl F}{\pl
 t_1}\right)^2-\frac{n}{2}(\sigma_1-b_1), \nonumber \\  &&
\nonumber\\
 - {\cal B}_1 F  &=& \left(\sigma_3
 \frac{\pl}{\pl t_3}+b_0 \frac{\pl}{\pl t_2}
 -(\sigma_1-b_1)\frac{\pl}{\pl t_1}  +\sum_{i\geq 1} it_i (\frac{\pl}{\pl
t_{i+3}}
 -\frac{\pl}{\pl t_{i+1}})
 \right.   \nonumber \\&& \left.
  +\beta \frac{\pl^2}{\pl t_{1}\pl t_2}\right)F
  +\beta \frac{\pl F}{\pl t_{1}} \frac{\pl F}{\pl t_{2}},
 \nonumber\\
  - {\cal B}_2 F  &=& \left(\sigma_4
 \frac{\pl}{\pl t_4}+b_0 \frac{\pl}{\pl t_3}
 -(\sigma_2-b_1)\frac{\pl}{\pl t_2} +\sum_{i\geq 1} it_i (\frac{\pl}{\pl
t_{i+4}}
 -\frac{\pl}{\pl t_{i+2}})
 \right.   \nonumber \\
&&\hspace{-0.5cm} \left. +\frac{\beta}{2}(
\frac{\pl^2}{\pl t_{2}^2}-
  \frac{\pl^2}{\pl t_{1}^2}+
  2\frac{\pl^2}{\pl t_{1}\pl t_3})\right)F
   +\frac{\beta}{2}\left(( \frac{\pl F}{\pl t_{2}} )^2-
  ( \frac{\pl F}{\pl t_{1}} )^2
  +2\frac{\pl F}{\pl t_{1}}\frac{\pl F}{\pl t_{3}}\right)
.\nonumber  \eea
\enddemo

\section{Matrix integrals and associated integrable systems}

 2.1. {\it Hermitian matrix integrals and the Toda lattice}.
Given a weight $\rho(z)=e^{-V(z)}$ defined as in (0.0.1),
the inner-product
\be
\la
f,g\ra_t=\int_{E}f(z)g(z)\rho_t(z)dz,~~~\mbox{with}~\rho_t:=e^{\sum^\infty_1
t_iz^i} \rho(z),\hskip.5in \spn{2.1.1}\ee leads to a moment matrix
\be
m_{n}(t)=(\mu_{ij}(t))_{0\leq i,j<n}=(\la
z^i,z^j\ra_t)_{0\leq i,j<n},\spn{2.1.2} \ee which is a {\em
H\"ankel matrix}\footnote{ H\"ankel means $\mu_{ij}$
depends on $i+j$ only.}, thus symmetric. H\"ankel is
tantamount to $\Lambda m_{\iy}=m_{\iy}\Lambda^{\top}$.
The semi-infinite moment matrix $m_{\iy}$ evolves in
$t$ according to the equations
\be
\frac{\pl\mu_{ij}}{\pl
t_{k}}=\mu_{i+k,j},\mbox{\,\,and thus\,\,} \frac{\pl
m_{\iy}}{\pl
t_{k}}=\Lambda^km_{\iy}~~~~~\left(\begin{array}{l}
\mbox{commuting}\\ \mbox{vector fields}
\end{array} 
\right).\qquad \spn{2.1.3}\ee
 Another important ingredient is the
factorization of $m_{\iy}$ into a lower- times an
upper-triangular matrix\footnote{This factorization is
possible for those $t$'s for which
$
\tau_n(t):=\det m_n(t)\neq 0 $ for all $n>0$.} 
$$ m_{\iy}(t)=S(t)^{-1}S(t)^{\top -1},
$$
where 
$S(t)$ is lower-triangular with nonzero diagonal elements.

\proclaim{Theorem} The vector $\tau(t)=(\tau_n(t))_{n\geq
0}${\rm ,} with
 \be
  \tau_n(t): =\det m_n(t)
 =\frac{1}{n!}\int_{E^n}\Delta^{2}_n(z)\prod^n_{k=1}\rho_t(z_k)dz_k
\spn{2.1.4}\ee
satisfies:
 \medbreak
 {\rm (i)}  {\rm Virasoro constraints}  {\rm (1.1.4)}
for $\beta =2${\rm ,}
 \be 
\left(-\sum_1^{2r} c_i^{k+1}f(c_i)\frac{\pl}{\pl c_i}+
\sum_{i\geq 0}\left( a_i~ \BJ_{k+i}^{(2)}-b_i ~
 \BJ_{k+i+1}^{(1)}\right) \right)\tau = 
0 \spn{2.1.5}
 \ee 
\medbreak
  {\rm (ii)}  the  {\rm KP-hierarchy}\footnote{for the
customary Hirota symbol $p(\pl_t)f\circ g:=
p(\frac{\pl}{\pl y})f(t+y)g(t-y) \Bigl|_{y=0}${\rm .} The
$p_{\ell}$\/{\rm '}\/s are the elementary Schur polynomials
$e^{\sum^{\iy}_{1}t_iz^i}:=\sum_{i\geq 0} p_i(t_1, t_2,\ldots)z^i$
and $p_{\ell}(\tilde \pl):=p_{\ell}(\frac{\pl}{\pl
t_1},\frac{1}{2}\frac{\pl}{\pl t_2},\ldots).$}
  $$
\left(p_{k+4}(\tilde\pl)-\frac{1}{2}\frac{\pl^2}{\pl
t_1\pl t_{k+3}}\right)\tau_n \circ\tau_n=0, $$
\phantom{(ii) } \hglue17pt of
which the first equation reads\/{\rm :} $$
\left(\left(\frac{\pl}{\pl t_1}
\right)^4+3\left(\frac{\pl}{\pl
t_2}\right)^2-4\frac{\pl^2}{\pl t_1 \pl
t_3}\right)\log\tau_n+6\left(\frac{\pl^2}{\pl
t^2_1}\log\tau_n \right)^2=0, $$ \hfill
$k=0,1,2,\ldots$

 {\rm (iii)}  The  {\rm standard Toda lattice}{\rm ;} i{\rm .}e{\rm .,} the tridiagonal
matrix
 \be
  L(t):=S(t)\Lb S(t)^{-1}=
\left(\begin{tabular}{lllll} $\frac{\pl}{\pl t_1}\log
\frac{\tau_1}{\tau_0}$ &
 $ \left(\frac{\tau_{0}\tau_{2}}{\tau_{1}^2}\right)^{1/2}$
 & ~~ $0$      &    \\
$\left(\frac{\tau_{0}\tau_{2}}{\tau_{1}^2}\right)^{1/2}$&
$\frac{\pl}{\pl t_1}\log \frac{\tau_2}{\tau_1}$ &
$\left(\frac{\tau_{1}\tau_{3}}{\tau_{2}^2}\right)^{1/2}$&
\\~~~$0$&$\left(\frac{\tau_{1}\tau_{3}}{\tau_{2}^2}\right)^{1/2}$
&
 $\frac{\pl}{\pl t_1}\log \frac{\tau_3}{\tau_2}$&
   \\
 & &  &      $\ddots$\\
\end{tabular}
\right) \spn{2.1.6}
 \ee
\phantom{(ii) } \hglue17pt  satisfies the commuting equations\footnote{$()_{\scriptscriptstyle\frak
s}$ means\/{\rm :} take the skew{\rm -}symmetric part of $()$ in the
decomposition {\rm ``}\/skew\/{\rm -}\/symmetric\/{\rm "} $+$
{\rm ``}\/lower\/{\rm -}\/triangular\/{\rm ."}}
 \be\frac{\pl L}{\pl
t_k}=\left[\frac{1}{2}(L^k)_{\frak s},L\right]. \spn{2.1.7}
 \ee

\ha {\rm (iv)}   {\rm Orthogonal polynomials}\/{\rm :}\/
 The $n^{\rm th}$ degree polynomials $p_n(t;z)$ in $z${\rm ,}
depending on $t \in \BC^{\iy}${\rm ,} orthonormal with
respect to the $t$\/{\rm -}\/dependent inner product {\rm (2.1.1)} $$ \la
p_k(t;z),p_{\ell}(t;z)\ra =\delta_{k\ell} $$
 are eigenvectors of $L${\rm ,} i.e.{\rm ,}
 $(L(t)p(t;z))_n=zp_n(t;z)
 ,~~{n\geq 0}${\rm ,} and
 enjoy the following representations
 \bean
p_n(t;z):=(S(t)\chi(z))_n&=&\frac{1}{\sqrt{\tau_{n}
(t)\tau_{n+1}(t)}}\det\left(
\begin{array}{lll|l}
  & & &1\\
  &m_n& &z\\
  & & &\vdots\\
 \hline
 \mu_{n,0}&\ldots&\mu_{n,n-1}&z^n
\end{array}
\right)\\
\navs
&=&z^nh_{n}^{-1/2}\frac{\tau_n(t-[z^{-1}])}{\tau_{n}
(t)},
~~~~h_{n}:=\frac{\tau_{n+1}(t)}{\tau_{n}(t)}.\eean

\ha \phantom{\rm (iv)} The functions
  $
 q_n(t;z):=
 z\int_{\BR^n}\frac{p_n(t;u)}{z-u}\rho_t(u)du
  $ are {\rm ``}\/dual eigenvectors\/{\rm "} of $L${\rm ,} i.e.{\rm ,} $(L(t)q(t;z))_n=zq_n(t;z)
 ,~~n\geq 1${\rm ,} and have
the following
 $\tau$\/{\rm -}\/function representation\/{\rm : (}\/see the remark at the
 end of this section\/{\rm )}\/
 
\bea
q_n(t;z):=z\int_{\BR^n}\frac{p_n(t;u)}{z-u}\rho_t(u)du
&=&\left(S^{\top -1}(t)\chi(z^{-1})\right)_n
\spn{2.1.8}\\
&=&\left(S(t)m_{\iy}(t)\chi(z^{-1})\right)_n\nonumber\\
&=&z^{-n}h_n^{-1/2}
\frac{\tau_{n+1}(t+[z^{-1}])}{\tau_n(t)}.\nonumber 
\eea

\ha {\rm (v)}  {\rm Bilinear relations:} for all
$n,m\geq 0${\rm ,} and $a,b \in \BC^{\iy}${\rm ,} such that
$a-b=t-t^{\prime}${\rm ,}

 \bea &
&\oint_{z=\iy}\tau_{n}(t-[z^{-1}])\tau_{m+1}(t'+[z^{-1}])
e^{\sum_1^{\iy}a_iz^i} z^{n-m-1}\frac{dz}{2\pi
i}\spn{2.1.9}\\
&&\quad=\oint_{z=0}\tau_{n+1}(t+[z])\tau_{m}(t'-[z])
e^{\sum_1^{\iy}b_iz^{-i}}z^{n-m-1}\frac{dz}{2\pi
i}.\nonumber  \eea
\endproclaim

  In the case $\beta=2$, the Virasoro expressions take on a particularly
  elegant form, namely for ${n \geq 0}$,
  \bean \BJ_{k,n}^{(2)}(t)&=&\sum_{i+j=k}
   :~ \BJ_{i,n}^{(1)}(t) ~
\BJ_{j,n}^{(1)}(t):~=~  J_k^{(2)}(t) + 2n J_k^{(1)}(t)
+ n^2\dt_{0k}\\ \navs
 \BJ_{k,n}^{(1)}(t) &=& ~  J_k^{(1)}(t)+
 n\dt_{0k} ,
  \eean
 with\footnote{The expression $J_k^{(1)}=0$ for $k=0$. }
\begin{eqnarray}
J_k^{(1)}&=&\frac{\pl}{\pl t_k}+\frac{1}{2}(-k)t_{-k},\spn{2.1.10}\\
J^{(2)}_{k}&=&\sum_{i+j=k}\frac{\pl^2}{\pl
 t_{i}\pl t_{j}}+ \sum_{-i+j=k}it_{i}\frac{\pl}{\pl
 t_{j}}+\frac{1}{4}\sum_{-i-j=k}it_{i}jt_{j}. \nn
 \end{eqnarray}
Statement (i) is already contained in Theorem 1.1,
whereas the other statements can be found in
\cite{AvM1}, \cite{AvM2}, and \cite{AvM6}. Notice that the standard Toda
lattice is a reduction of the semi-infinite 2-Toda
lattice, where $\tau_n(t,s)=\tau_n(t-s)$. The 2-Toda
lattice arises in the context of a factorization of a
generic semi-infinite matrix $m_{\iy}(t,s)$,
satisfying the simple equations $\frac{\pl
m_{\iy}}{\pl t_k}=\Lb^km_{\iy}, \frac{\pl m_{\iy}}{\pl
s_k}=-m_{\iy} \Lb^{\top k},$ whereas the standard Toda
lattice is related to the same factorization of
$m_{\iy}(t,s)$, but where $m_{\iy}(t,s)$ is H\"ankel
(i.e., $\Lb m_{\iy}=m_{\iy}\Lb^{\top}$).

\demo{{R}emark} {\it The vectors $p$ and $q$ are eigenvectors
of $L$}. Indeed, remembering
$\chi(z)=(1,z,z^2,...)^{\top}$, we have
 $$ \Lb \chi(z)=z\chi(z)~~\mbox{and}~~
  \Lb^{\top}
  \chi(z^{-1})=z\chi(z^{-1})-ze_1,~~\mbox{with}~e_1=(1,0,0,...)^{\top}.
  $$
  Therefore, $p(z)=S\chi(z)$ and $q(z)=S^{\top
  -1}\chi(z^{-1})$ are eigenvectors, in the sense
  \bean
  Lp&=&S\Lb S^{-1}S\chi(z)=zS\chi(z)=zp  ,\\ \navs
  L^{\top}q&=& S^{\top -1}\Lb^{\top} S^{\top}
   S^{\top -1} \chi(z^{-1})\\ \navs
&=&zS^{\top -1}\chi(z^{-1})
   -zS^{\top -1}e_1=zq-zS^{\top -1}e_1 .
   \eean
Then, using $L=L^{\top}$, one is lead to
$$((L-zI)p)_n=0,~~\mbox{for}~ n\geq 0~~~\mbox{and}~~~
 ((L-zI)q)_n=0,~~\mbox{for}~~ n\geq 1.$$
\enddemo

\demo{{\rm 2.2.} Symmetric\/{\rm /}\/symplectic
 matrix integrals and the Pfaff lattice}
Consider an inner-product, with a skew-symmetric
weight $\rho(y,z)$, \be \la
f,g\ra_t=\int\!\int_{\BR^2}f(y)g(z)e^{\sum^\infty_1
t_i(y^i+z^i)}\rho(y,z)dy\,dz,\mbox{\,\,with
$\rho(z,y)=-\rho(y,z)$}. \spn{2.2.1}\ee 
Then, since $$\la
f,g\ra_t=-\la g,f\ra_t $$ the (semi-infinite) moment
matrix, depending on $t=(t_1,t_2,\ldots)$, $$
m_{n}(t)=(\mu_{ij}(t))_{0\leq i,j\leq n-1}=(\la
y^i,z^j\ra_t)_{0\leq i,j\leq n-1} $$ is skew-symmetric
and the semi-infinite matrix $m_{\iy}$ evolves in $t$
according to the {\em commuting vector fields}
\be
\frac{\pl\mu_{ij}}{\pl
t_k}=\mu_{i+k,j}+\mu_{i,j+k},\mbox{\,\,i.e.,}~~
\frac{\pl m_{\iy}}{\pl
t_k}=\Lb^km_{\iy}+m_{\iy}\Lb^{\top k}. \hskip.5in \spn{2.2.2}\ee
It is well known that the determinant of an odd
skew-symmetric matrix equals 0, whereas the
determinant of an even skew-symmetric matrix is the
square of a polynomial in the entries, the Pfaffian,
with a sign specified below. So
 \bean
\det(m_{2n-1}(t))&=&0\\ (\det
m_{2n}(t))^{1/2}&=&pf(m_{2n}(t))=\frac{1}{n!}(dx_{0}\wedge
dx_1\wedge \ldots\wedge dx_{2n-1})^{-1}\\ &
&\hspace{4cm}\left(\sum_{0\leq i<j\leq
2n-1}\mu_{ij}(t)dx_i\wedge dx_j\right)^n. \eean
 Define now the {\em Pfaffian $\tau$-functions}:
 \be
 \tau_{2n}(t):=pf~ m_{2n}(t), \spn{2.2.3}
 \ee
and the semi-infinite skew-symmetric matrix, $0$
everywhere, except for the $2\times 2 $ blocks, along
the diagonal:
  \be J:=\left(
\begin{array}{cc@{}c@{}cc}
 &\fbox{$\begin{array}{cc} 0 & 1 \\ -1 & 0 \end{array}$} && & \\
 && \fbox{$\begin{array}{cc} 0 & 1 \\ -1 & 0 \end{array}$} &&\\
 &&& \fbox{$\begin{array}{cc} 0 & 1 \\ -1 & 0 \end{array}$} & \\
 &&&& \ddots
 \end{array}
 \right) \mbox{,  with $J^2=-I$}.\quad \spn{2.2.4}
\ee
 Since $m_{\iy}$ is skew-symmetric, $m_{\iy}$ does
not admit a Borel factorization in the standard sense,
but $m_{\iy}$ admits a unique factorization, with the
matrix $J$ inserted (see \cite{AHV}):
 $$ m_{\iy}(t)=Q^{-1}(t)J\,Q^{\top
-1}(t),
 $$  
 where
\be 
 Q(t) = \left(
\begin{array}{c@{}c@{}cc}
\ddots &&&0 \\
&&0& \\
 & \fbox{$\begin{array}{cc}
 Q_{2n,2n} & 0 \\ 0 & Q_{2n,2n} \end{array}$} &&\\
 &*& \fbox{$\begin{array}{cc}
  Q_{2n+2,2n+2} & 0 \\ 0 & Q_{2n+2,2n+2} \end{array}$} & \\
 &&& \ddots
 \end{array}
 \right) ~\in K. \spn{2.2.5} 
 \ee
\medbreak\noindent 
 $K$ is the group of lower-triangular invertible
matrices of the form above, with Lie algebra
${\frak k}$ of matrices of precisely the same  form.
In this problem, the Lie algebra splitting of
semi-infinite matrices is given by \be gl(\iy)={\frak
k}\oplus{\frak n}\left\{
\begin{array}{l}
{\frak k}=\{\mbox{lower-triangular matrices of the
form (2.2.5)}\}\\ \navs {\frak n}=sp(\iy)=\{\mbox{$a$ such that
$Ja^{\top}J=a$}\},
\end{array}
\right. 
\spn{2.2.6}\ee with unique decomposition ($a_{\pm}$
refers to projection onto strictly upper- (strictly
lower) triangular matrices, with all $2\times 2$
diagonal blocks equal to zero)
 \bea
  a&=&(a)_{\frak k}
+(a)_{\frak n} \spn{2.2.7}
 \\ \navs
&=&\left((a_--J(a_+)^{\top}J)+\frac{1}{2}
(a_0-J(a_0)^{\top}J)\right)  \nonumber
 \\ \navs &&~~~~~~~~~+
\left((a_++J(a_+)^{\top}J)+\frac{1}{2}(a_0+J(a_0)^{\top}
J)\right). \nn
 \eea

 Considering as a
special 
skew-symmetric weight (2.2.1),
\be
 \rho(y,z):= 2 D^{\alpha}\delta (y-z)\tilde \rho (y)
  \tilde \rho (z)~, \mbox{with}~\alpha=\mp 1
,~~~\tilde \rho(y)=e^{-\tilde V(y)} ,\spn{2.2.8} \ee
 the inner-product (2.2.1)
 becomes\footnote{$\vr(y)= sign (y)$,  and $\{f,g\}:=
 f'g-fg'$. Also notice that
$\vr^{\prime}=2\delta(x)$.} (see \cite{ASV3})
  \bean \la
f,g\ra_t&=&\int\!\int_{\BR^2}f(y)g(z)e^{\sum
t_i(y^i+z^i)}2 D^{\alpha}\delta (y-z)\tilde \rho (y)
  \tilde \rho (z)dy\,dz  \\ &&\\ \navs
 &=&\left\{\begin{array}{ll}
\displaystyle{\int\!\!\!\int_{\BR^2}}
f(y)g(z)e^{\sum_1^{\iy}t_i(y^i+z^i)}\vr(y-z)
 \tilde\rho(y)\tilde\rho(z)dy\,dz
, &\mbox{\,\,for\,\,}\alpha =-1\\
 & \\
\displaystyle{\int_{\BR}}\{f,g\}(y)e^{\sum_1^{\iy}2t_i
y^i}
 \tilde\rho(y)^2dy,&
\mbox{\,\,for\,\,}\alpha =+1,
\end{array}
\right. \eean
 and (see \cite{Mehta}, \cite{AvM5})
 \bea
 &&\spn{2.2.9}\\
  \lefteqn{pf\left(\la
y^i,z^j\ra_t\right)_{0\leq i,j\leq 2n-1}}\nn\\ \navs
&&\quad =\left\{\begin{array}{ll}
\displaystyle{\frac{1}{(2n)!}\int_{\BR^{2n}}|\Dt_{2n}(z)|
\prod^{2n}_{k=1}e^{\sum_1^{\iy}t_iz_k^i}\tilde\rho(z_k)dz_k}\\ \navs
\hskip1in{\displaystyle  =
 \frac{1}{(2n)!}\int_{{\cal
S}_{2n}} e^{\Tr (-\tilde V(X)+\sum t_iX^i)}dX},&\enspace \mbox{for\,\,}\al=-1,\\ \navs
\displaystyle{\frac{1}{n!}\int_{\BR^{n}}|\Dt_{n}(z)|^4
 \prod^{n}_{k=1}e^{\sum_1^{\iy}2t_iz_k^i}\tilde\rho^2(z_k)dz_k}\\ \navs
\hskip1in {\displaystyle=\frac{1}{n!}
  \int_{{\cal T}_{2n}} e^{\Tr
(-2\tilde V(X)+\sum 2t_iX^i)}dX},&\enspace
\mbox{for\,\,}\al=+1.
\end{array}
\right. \nonumber \eea
Setting
  $$ \left\{
\begin{array}{ll} \tilde \rho(z)=\rho(z)I_E(z) &
\mbox{for}~\alpha=-1\\ \navs
 \tilde
\rho(z)=\rho^{1/2}(z)I_E(z),~t\mapsto t/2 &
\mbox{for}~\alpha=+1
 \end{array}
 \right.
  $$
in the identities (2.2.9), we are led to the identities
between integrals and Pfaffians, which are spelled out in
Theorem 2.2:
\enddemo

\phantom{changes}

\proclaim{Theorem} The integrals $I_n(t,c)${\rm ,}
\bean
 I_n& =&\int_{E^n}|\Dt_n(z)|^{\beta}\prod_{k=1}^n
\left(e^{\sum_1^{\iy}t_i z_k^i}\rho(z_k)dz_k\right)
\\ \navs
&=&\left\{
\begin{array}{l}
\displaystyle{n!pf\left( \int\!\!\!\int_{E^2} y^iz^j
 \varepsilon(y-z)e^{\sum_1^{\iy}t_k (y^k+z^k)}\rho(y)
 \rho(z) dydz\right)_{0\leq i,j\leq n-1}
 }\\ \navs
\hskip1in {\displaystyle=n!\tau_n(t,c),} \hskip1.15in  n~\mbox{even{\rm ,} for}~\beta=1
 \\  \\\displaystyle{ n! pf~\left( \int_E\{y^i,y^j\}
 e^{\sum_1^{\iy}t_k y^k}\rho(y)dy
  \right)_{0\leq i,j\leq 2n-1}=n!
\tau_{2n}(t/2,c),}\\ \navs \hskip2.9in
 n~\mbox{arbitrary{\rm ,}
 for}~\beta=4 
\end{array}
\right. \eean
  and the $\tau_n(t,c)$\/{\rm '}\/s above satisfy the following
 equations\/{\rm :}

\bigbreak
 \ha {\rm (i)}  The
 {\rm Virasoro constraints}\footnote{here the $a_i$'s and $b_i$'s are defined in the
  usual way, in terms of $\rho(z)$; namely,
  $-\frac{\rho'}{\rho}=\frac{\sum_{\phantom{1}}\!\! b_i z^i}{\sum^{\phantom{1}}\!\! a_i z^i}$.
  } {\rm (1.1.4)} for $\beta =1,4$,

 \be
\left(-\sum_1^{2r} c_i^{k+1}f(c_i)\frac{\pl}{\pl c_i}+
\sum_{i\geq 0}\left( a_i~
{}^{\beta}\BJ_{k+i,n}^{(2)}-b_i ~
 {}^{\beta}\BJ_{k+i+1,n}^{(1)}\right) \right)I_n =
0\hskip.25in \spn{2.2.10}
 \ee

 {\rm (ii)}  The {\rm Pfaff-KP hierarchy:} {\rm (}\/see footnote {\rm 6)}
\be
\left(p_{k+4}(\tilde\pl)-\frac{1}{2}\frac{\pl^2}{\pl
t_1\pl t_{k+3}}\right)\tau_{n}\circ\tau_{n}=p_k(\tilde
\pl)~\tau_{n+2}\circ\tau_{n-2}\qquad\qquad \spn{2.2.11} \ee \hfill $n
~\mbox{even},~ k=0,1,2,...~.$

\ha\phantom{\rm (ii)} of which the first equation reads
 
$$ \left(\left(\frac{\pl}{\pl t_1}
\right)^4+3\left(\frac{\pl}{\pl
t_2}\right)^2-4\frac{\pl^2}{\pl t_1 \pl
t_3}\right)\log\tau_n+6\left(\frac{\pl^2}{\pl
t^2_1}\log\tau_n
\right)^2=12\frac{\tau_{n-2}\tau_{n+2}}{\tau_{n}^2},
$$ \hfill $n$ even{\rm .}

\medbreak \ha {\rm (iii)}  The {\rm Pfaff lattice:} The
time\/{\rm -}\/dependent matrix

 \be
L(t)=Q(t)\Lb Q(t)^{-1} \spn{2.2.12}
 \ee

\ha \phantom{\rm (iii)} satisfies the
 Hamiltonian commuting equations{\rm ,} given by the
 Adler\/{\rm -}\/Kostant\/{\rm -}\/Symes splitting theorem{\rm ,} applied to the
 splitting $gl(\iy)={\frak
k}\oplus{\frak n}${\rm ,} as in {\rm (2.2.6)} and {\rm (2.2.7),}
$$\frac{\pl L}{\pl t_{i}}=[-(L^i)_{\frak
k},L],~~~~~~~\mbox{\rm (Pfaff lattice) }$$

\ha {\rm (iv)}  {\rm Skew-orthogonal polynomials:}
 The vector of time\/{\rm -}\/dependent polynomials
$q(t;z):=(q_n(t;z))_{n\geq 0}=Q(t)\chi(z)$ in $z$
 satisfy the eigenvalue problem
\be
L(t)q(t,z)=zq(t,z)\spn{2.2.13} \ee 

\ha \phantom{\rm (iv)} and
 enjoy the following representations\/{\rm :}
 \bean
q_{2n}(t;z)&=&z^{2n}h_{2n}^{-1/2}
 \frac{\tau_{2n}(t-[z^{-1}])}{\tau_{2n}(t)} ~~,~~~~
 h_{2n}=\frac{\tau_{2n+2}(t)}{\tau_{2n}(t)} \\ \navs
q_{2n+1}(t;z)&=&z^{2n}h_{2n}^{-1/2}\frac{1}{\tau_{2n}(t)}\left(z+\frac{\pl}{\pl
t_1}\right)
{\tau_{2n}(t-[z^{-1}])}. \eean

\ha \phantom{\rm (iv)} They are
skew-orthogonal polynomials in $z${\rm ;} i.e.{\rm ,} $$ \la
q_i(t;z),q_j(t;z)\ra_t=J_{ij} .$$

\ha {\rm (v)}   {\rm The bilinear identities:}
For all $n,m\geq 0${\rm ,} the $\tau_{2n}$\/{\rm '}\/s satisfy the
following bilinear identity

 \bea &
&\oint_{z=\iy}\tau_{2n}(t-[z^{-1}])\tau_{2m+2}(t'+[z^{-1}])
e^{\sum^\infty_1(t_i-t'_i)z^i} z^{2n-2m-2}\frac{dz}{2\pi
i}\spn{2.2.14}\\
&&\quad+\oint_{z=0}\tau_{2n+2}(t+[z])\tau_{2m}(t'-[z])
e^{\sum^\infty_1(t'_i-t_i)z^{-i}}z^{2n-2m}\frac{dz}{2\pi
i}=0.\nonumber  \eea
 
\endproclaim

Note that (2.2.10) is a consequence of Theorem 1.1, while
items (ii) to (v) are shown in \cite{AvM5}, \cite{AHV}. (See
\cite{ASV3} for the Pfaff lattice, viewed as a
reduction of the 2-Toda lattice.) A semi-infinite
matrix $m_{\iy}(t,s)$, satisfying $\frac{\pl m_{\iy}}{\pl
s_k}=\Lb^km_{\iy}, \frac{\pl
m_{\iy}}{\pl t_k}=-m_{\iy} \Lb^{\top k},$ leads to the
semi-infinite 2-Toda lattice. When the initial condition $m_{\iy}(0,0)$ is
skew-symmetric, then $m_\iy(t,-t)$ remains skew-symmetric in time and 
$\tau_n(t)=(\tau_n(t,-t))^{1/2}=pf m_{n}(t,-t)$ is a
Pfaff lattice $\tau$-function.

 \section{Expressing $t$-partials in terms of
boundary-partials}
 
\demo{{\rm 3.1.} Gaussian and Laguerre ensembles}
Given first-order linear operators ${\cal D}_1,{\cal
D}_2,{\cal D}_3$ in $c=(c_1,...,c_{2r})\in\BR^{2r}$
and a function $F(t,c)$, with $t\in\BC^{\iy}$,
satisfying the following partial differential
equations in $t$ and $c$:
\be
{\cal D}_kF=\frac{\pl F}{\pl t_k}+\sum_{-1\leq
j<k}\ga_{kj}V_j(F)+\ga_k+\delta_kt_1,\quad k=1,2,3,\ldots , \hskip.25in  \spn{3.1.1}
\ee with $V_j(F)$ nonlinear differential operators in
$t_i$ of which the first few are given here:
\be
V_j(F)=\sum_{i,i+j\geq 1}it_i\frac{\pl F}{\pl
t_{i+j}}+\frac{\beta}{2}\delta_{2,j}\left(\frac{\pl^2
F}{\pl t_1^2}+\left(\frac{\pl F}{\pl
t_1}\right)^2\right),\quad -1\leq j\leq 2.\quad \spn{3.1.2} \ee
 In (3.1.1) and (3.1.2),
$\beta >0,\gamma_{kj},\gamma_k, \delta_k$ are
arbitrary parameters; also $\delta_{2j}=0$ for $j\neq
2$ and $\delta_{2j}=1$ for $j=2$. The claim is that the equations
(3.1.1) enable one to express all partial derivatives,
 \be
 \left.\frac{\pl^{i_1+...+i_k} F(t,c)}{\pl t_1^{i_1}
 ...\pl t_k^{i_k}}\right|_{\cal L},~\mbox{along}~{\cal
 L}:=\{\mbox{all $t_i=0, ~c=(c_1,...,c_{2r})$
 arbitrary}\}, \spn{3.1.3}
 \ee
 uniquely in terms of polynomials in
 ${\cal D}_{j_1}...{\cal D}_{j_r}F(0,c).$ 
 Indeed, the method consists of expressing
$\displaystyle{\frac{\pl F}{\pl t_k}\Biggl|_{t=0_{\phantom{|}}}}$ in
terms of $\displaystyle{{\cal D}_kf\Biggl|_{t=0}}$,
using (3.1.1). Second derivatives are obtained by
acting on ${\cal D}_kF$ with ${\cal D}_{\ell}$, by
noting that ${\cal D}_{\ell}$ commutes with all
$t$-derivatives, by using the equation for ${\cal
D}_{\ell}F$, and by setting in the end $t=0$: \bea
{\cal D}_{\ell}{\cal D}_kF&=&{\cal D}_{\ell}\frac{\pl
F}{\pl t_k} +\sum_{-1\leq j<k}\ga_{kj}{\cal
D}_{\ell}(V_j(F))\nonumber\\
\navs &=&\left(\frac{\pl }{\pl
t_k} +\sum_{-1\leq j<k}\ga_{kj}V_j\right){\cal
D}_{\ell}(F),\quad\mbox{provided $V_j(F)$ does
not}\nonumber\\ & &\hspace{5.5cm}\mbox{contain
nonlinear terms}\nonumber\\ \navs &=&\left(\frac{\pl }{\pl
t_k} +\sum_{-1\leq
j<k}\ga_{kj}V_j\right)\left(\frac{\pl F}{\pl t_{\ell}}
+\sum_{-1\leq j<\ell}\ga_{\ell
j}V_j(F)+\dt_{\ell}t_1\right)\nonumber\\ \navs
&=&\frac{\pl^2F}{\pl t_k\pl
t_{\ell}}+\mbox{\,lower-weight terms.}\nonumber \eea
When the nonlinear term is present, it is taken care of
as follows:
\begin{eqnarray*}
{\cal D}_{\ell}\left( \frac{\pl F}{\pl t_1} \right)^2
&=&2\frac{\pl F}{\pl t_1} {\cal D}_{\ell} \frac{\pl
F}{\pl t_1}\\ \navs &=& 2\frac{\pl F}{\pl t_1} \frac{\pl
}{\pl t_1} {\cal D}_{\ell}F\\ \navs  &=&2\frac{\pl F}{\pl
t_1} \frac{\pl }{\pl t_1} \left(  \frac{\pl F}{\pl
t_{\ell}}+\sum_{-1\leq
j<{\ell}}\ga_{{\ell}j}V_j(F)+\ga_{\ell}+\delta_{\ell}t_1
\right)
;\end{eqnarray*}
 higher derivatives are obtained in
the same way. Explicit expressions for only a few
partials, useful in the next subsection, will be given
here:
\begin{eqnarray}
&& \spn{3.1.4}\\
\frac{\pl F}{\pl t_1}\Biggl|_{\cal L}&=&{\cal D}
_1F-\ga_1, \nn \\ \navs  \frac{\pl^2 F}{\pl
t^2_1}\Biggl|_{\cal L}&=&\left({\cal
D}_1^2-\ga_{10}{\cal
D}_1\right)F+\ga_{10}\ga_1-\dt_1,\nonumber\\ \navs 
\frac{\pl^3 F}{\pl t^3_1}\Biggl|_{\cal
L}&=&\left({\cal D}_1^3-3\ga_{10}{\cal D}_1^2+
2\ga_{10}^2{\cal D}_1\right)F+2\ga_{10}
(\dt_1-\ga_1\ga_{10}),\nonumber\\ \navs  \frac{\pl^4 F}{\pl
t^4_1}\Biggl|_{\cal L}&=&\left({\cal
D}_1^4-6\ga_{10}{\cal D}_1^3+11\ga_{10}^2{\cal
D}_1^2-6\ga_{10}^3 {\cal
D}_1\right)F-6\ga_{10}^2(\dt_1-\ga_1\ga_{10}),\nonumber\\ \navs 
\frac{\pl F}{\pl t_2}\Biggl|_{\cal L}&=&{\cal
D}_2F-\ga_2,\nonumber\\ \navs  \frac{\pl^2 F}{\pl
t^2_2}\Biggl|_{\cal L}&=&\biggl({\cal
D}_2^2-2\ga_{20}{\cal D}_2+\beta\ga_{21} \ga_{32}{\cal
D}^2_1\nn\\ \navs  &&\phantom{\Big(}\,-((2\ga_1
+\ga_{10})\ga_{21}\ga_{32}\beta+2\ga_{2,-1}) {\cal
D}_1 -\ 2\ga_{21}{\cal D}_3\biggr)F\nn \\ \navs  &&
+\ \beta\ga_{21}\ga_{32}({\cal D}_1F)^2
+ \beta\ga_{21}\ga_{32}(\ga^2_1+\ga_{10}
\ga_1-\dt_1)\nn \\ \navs  &&
+\ 2(\ga_{21}\ga_3+\ga_{20}\ga_2+\ga_1\ga_{2,-1}),\nonumber\\ \navs 
\frac{\pl F}{\pl t_3}\Biggl|_{\cal L}&=&\left({\cal
D}_3- \frac{\beta}{2}\ga_{32}{\cal
D}_1^2+\frac{\beta}{2} \ga_{32}(2\ga_1+\ga_{10}){\cal
D}_1\right)F-\frac{\beta}{2}\ga_{32}({\cal
D}_1F)^2\nonumber\\ \navs  &
&\phantom{\Big(}\, +\ \frac{\beta}{2}\ga_{32}(\dt_1-\ga_1\ga_{10}-\ga^2_1)-\ga_3,\nonumber\\ \navs 
\frac{\pl^2 F}{\pl t_1\pl t_3}\Biggl|_{\cal
L}&=&\biggl({\cal D}_1{\cal
D}_3-\frac{\beta}{2}\ga_{32}{\cal D}_1^3
+\beta\ga_{32}( \ga_1+2\ga_{10}){\cal
D}_1^2\nonumber\\ \navs 
& &\phantom{\Big(}\, -\ \frac{3\beta}{2}
\ga_{10}\ga_{32}(2\ga_1+\ga_{10}){\cal D}_1-  3\ga_{1,-1}{\cal D}_2-3\ga_{10}{\cal
D}_3\biggr)F
\nonumber\\ \navs  & &+\ \frac{3\beta}{2}\ga_{10}\ga_{32}({\cal
D}_1F)^2 -\beta\ga_{32}({\cal D}_1F)({\cal
D}_1^2F)\nonumber\\ \navs  & &+\ 
\frac{3}{2}(2\ga_{10}\ga_3+\beta\ga_{32}\ga_{10}(\ga^2_1+\ga_{10}\ga_1-\dt_1)+2\ga_{1,-1}
\ga_2).\nonumber
\end{eqnarray}
 \enddemo
 
\demo{{\rm 3.2.} Jacobi ensemble}

1. From the expressions (1.3.11), upon evaluating $ \left.{\cal
B}_{-1}F\right|_{t=0}, \left. {\cal
B}^2_{-1}F\right|_{t=0},\break \left.{\cal
B}_{0}F\right|_{t=0} ,$ one finds the following equations, both
sides of which are evaluated at $t=0$,
 \bean
 -{\cal B}_{-1}F&=&{ \sigma_1}{{\pl
}\over{\pl{t_1}}}F+b_0n,\\
\navs
 \frac{1}{\sigma_1}{\cal B}^2_{-1}F&=&\left(
\sigma_1{{\pl^{2} }\over
 {\pl { t_1}^{2}}}+{{\pl
 }\over{\pl { t_2}}}\right)F-n, \\
\navs
  -{\cal B}_{0}F&=&\left(b_0 \frac{\pl  }{\pl
 t_1}+ \sigma_2\frac{\pl}{\pl t_2}\right)F
  +\frac{\beta}{2}\left(\left(\frac{\pl }{\pl  t_1}\right)^2F
+ \left( \frac{\pl F}{\pl
 t_1}\right)^2\right)-\frac{n}{2}(\sigma_1-b_1). \\
 \eean
From these expressions, one extracts $$
 \left.\frac{\pl F}{\pl t_1}\right|_{t=0},
 \left.\frac{\pl^2 F}{\pl t_1^2}\right|_{t=0},
 \left.\frac{\pl F}{\pl t_2}\right|_{t=0},
$$
in terms of ${\cal B}_j^iF$.

\bigbreak
  2. From the expressions for $ \left.{\cal
B}^3_{-1}F\right|_{t=0}, \left.{\cal B}_{0} {\cal
B}_{-1}F\right|_{t=0}, \left.{\cal
B}_{1}F\right|_{t=0} $, namely
 \bean
  {\cal B}_{1}F&=&\left(-b_0\frac{\pl}{\pl
t_2}+(\sigma_1-b_1) \frac{\pl}{\pl t_1}-\sigma_3
\frac{\pl}{\pl t_3}\right)F-\beta
 \left(\frac{\pl^2 F}{\pl t_1\pl t_2}+\frac{\pl F}{\pl
t_1}\frac{\pl F}{\pl t_2}\right),\\
\navs
\frac{1}{\sigma_1}{\cal B}_{0}{\cal B}_{-1}F & =&
\left( \sigma_2  \frac{\pl^2}{\pl t_1 \pl
t_2}+\frac{\pl}{\pl t_3} -  \frac{\pl}{\pl t_1}
 +b_0  \frac{\pl^2}{\pl t_1^2}
 \right)F
 +\frac{\beta}{2} \left( \frac{\pl^3 F}{\pl t_1^3}
 +2  \frac{\pl F}{\pl t_1} \frac{\pl^2 F}{\pl t_1^2}
 \right),\\
\navs
 - \frac{1}{\sigma_1}{\cal B}^3_{-1}F&=&\left(
\sigma_1^2 \frac{\pl^3 }{\pl t_1^3} +3\sigma_1
\frac{\pl^2}{\pl t_1 \pl t_2} -2 \frac{\pl}{\pl t_1}
+2 \frac{\pl}{\pl t_3}\right)F,
 \eean
 one extracts $$
 \left.\frac{\pl F}{\pl t_3}\right|_{t=0},
 \left.\frac{\pl^2 F}{\pl t_1^3}\right|_{t=0},
 \left.\frac{\pl^2 F}{\pl t_1\pl t_2}\right|_{t=0}
$$
in terms of ${\cal B}_j^iF$, using the previous
extractions.
\bigbreak
  3. From the expressions for $ \left.{\cal
B}_{2}F\right|_{t=0}, \left.{\cal B}_{1} {\cal
B}_{-1}F\right|_{t=0}, \left.{\cal
B}^2_{0}F\right|_{t=0},
 \left.{\cal
B}_{0} {\cal B}^2_{-1}F\right|_{t=0}$,\break $\left. {\cal
B}^4_{-1}F\right|_{t=0}$,  namely, (where both sides are
evaluated at $t=0$)
\medbreak
\noindent (3.2.1)
\smallbreak
{\ninepoint 
 \begin{eqnarray*}
\noalign{\vskip-24pt}
  {\cal B}_{2}F&=& \left( -\sigma_4 \frac{\pl}{\pl t_4}
 -b_0 \frac{\pl}{\pl t_3}
 +(\sigma_2-b_1) \frac{\pl}{\pl t_2} +\frac{\beta}{2}
 \bigl(\frac{\pl^2}{\pl t_1^2}-\frac{\pl^2}{\pl t_2^2}
-2\frac{\pl^2}{\pl t_1 \pl t_3}
 \bigr)\right)F   \\ \navs
 &&+\ \beta \left( \Bigl(\frac{\pl F}{\pl t_1}\Bigr)^2
 - \Bigl(\frac{\pl F}{\pl t_2}\Bigr)^2
 -\frac{\pl F}{\pl t_1}\frac{\pl F}{\pl t_3} \right) , \nn \\
\navs
\frac{1}{\sigma_1}{\cal B}_{1}{\cal B}_{-1}F & =&
\left(\frac{\pl }{\pl t_4}-\frac{\pl }{\pl t_2}+b_0
\frac{\pl^2}{\pl t_1 \pl t_2}+\sigma_3
\frac{\pl^2}{\pl t_1 \pl t_3} -(\sigma_1-b_1)
\frac{\pl^2}{\pl t_1 ^2}+\beta \frac{\pl^3}{\pl t_1^2
\pl t_2}\right)F \nn \\ \navs &&~~~+\  \beta\left(\frac{\pl^2
F}{\pl t_1^2}\frac{\pl F}{\pl t_2}+ \frac{\pl F}{\pl
t_1}\frac{\pl^2 F}{\pl t_1 \pl t_2} \right) ,
  \\ \navs
  {\cal B}^2_{0}F&=&\left(b_0\frac{\pl}{\pl t_1}
+\sigma_2\frac{\pl}{\pl
t_2}+\frac{\beta}{2}\frac{\pl^2}{\pl
t_1^2}+\beta\frac{\pl}{\pl t_1}F\frac{\pl}{\pl
t_1}\right)\nn \\ \navs
 &&\left(b_0\frac{\pl F}{\pl t_1}
+\sigma_2\frac{\pl F}{\pl t_2}+\sum^2_1 it_i(\frac{\pl
F }{\pl t_{i+2}} -\frac{\pl F}{\pl t_i})
+\frac{\beta}{2} \Bigl(\frac{\pl^2 F}{\pl t_1^2} +
\bigr(\frac{\pl F}{\pl t_1}\bigl)^2\Bigr)\right),
 \nn \\ \navs
  \frac{1}{\sigma_1} {\cal B}_{0}{\cal B}^2_{-1}F & =&-\frac{\pl}{\pl t_1}
  \left( \sigma_1\frac{\pl}{\pl t_1}+
  t_1\frac{\pl}{\pl t_2} \right) \nn \\ \navs
  &&\left(b_0\frac{\pl F}{\pl t_1}
+\sigma_2\frac{\pl F}{\pl t_2}+\sum^2_1 it_i(\frac{\pl
F }{\pl t_{i+2}} -\frac{\pl F}{\pl t_i})
+\frac{\beta}{2} \Bigl(\frac{\pl^2 F}{\pl t_1^2} +
\bigr(\frac{\pl F}{\pl t_1}\bigl)^2\Bigr)\right) ,\nn \\
\navs
{\cal B}^4_{-1}F & =&
 \sigma_1\frac{\pl}{\pl t_1}
  \left( \sigma_1\frac{\pl}{\pl t_1}+
  t_1\frac{\pl}{\pl t_2} \right)
  \left( \sigma_1\frac{\pl}{\pl t_1}+
  t_1\frac{\pl}{\pl t_2}+
  2t_2(\frac{\pl}{\pl t_3}-\frac{\pl}{\pl t_1})
  \right)\nn \\ \navs
 && \left( \sigma_1\frac{\pl F}{\pl t_1}+
  t_1\frac{\pl F}{\pl t_2}+\sum_{2}^3
  it_i(\frac{\pl F}{\pl t_{i+1}}-
  \frac{\pl F}{\pl t_{i-1}})+b_0n-nt_1 \right),
 \end{eqnarray*} 
}
\smallbreak\noindent 
   one extracts
 \be
 \left.\frac{\pl ^4 F}{\pl t_1^4}\right|_{t=0},
 \left.\frac{\pl F}{\pl t_4}\right|_{t=0},
 \left.\frac{\pl^3 F}{\pl t^2_1\pl t_2}\right|_{t=0},
 \left.\frac{\pl^2 F}{\pl t_1\pl t_3}\right|_{t=0},
 \left.\frac{\pl^2 F}{\pl t_2^2}\right|_{t=0}, \hskip.5in\spn{3.2.2}
\ee
 again in terms of ${\cal B}_j^iF$, using all the previous
extractions.

\enddemo

\demo{{\rm 3.3.} Evaluating the matrix integrals
 on the full range}
The denominators of the probabilities (0.0.4), for
$\beta=1,4$; namely:
 $$
I_n^{(\beta)}:=\left\{\begin{array}{l}
 \displaystyle{\int_{\BR^n}|\Dt_n(z)|^{\beta}\prod_{k=1}^n
e^{-b z_k^2}dz_k } \\
\navs
\displaystyle{ \int_{\BR_+^n}|\Dt_n(z)|^{\beta}\prod_{k=1}^n
z_k^{a} e^{-b z_k}dz_k} \\
\navs
\displaystyle{ \int_{[-1,1]^n}|\Dt_n(z)|^{\beta}\prod_{k=1}^n
(1-z_k)^{a}(1+z_k)^{b}dz_k} \\
 \end{array}\right.,
 $$
can be evaluated, using Selberg's integral (see Mehta
\cite[p.  340]{Mehta}): $$
I_n^{(\beta)}\left\{\begin{array}{l}
=\displaystyle{(2\pi)^{n/2}(2b)^{-n(\beta(n-1)+2)/4}
\prod^{n-1}_{j=0}
  \frac{ \Gamma((j+1)\beta /2 +1 )}
{\Gamma(\beta /2 +1)}}\\
\navs
 =\displaystyle{b^{-n(\beta(n-1)+2a+2)/2} \prod^{n-1}_{j=0} \frac{\Gamma(a+1+j\beta /2
) \Gamma((j+1)\beta /2 +1)} {\Gamma(\beta /2 +1)}}\\
\navs
=\displaystyle{2^{n(2a+2b+\beta(n-1)+2)/2}}\\
 \displaystyle{~~~~~~~~~~~. \prod^{n-1}_{j=0} \frac{\Gamma(a+j\beta /2
+1)\Gamma(b+j\beta /2+1) \Gamma((j+1)\beta /2 +1)}
{\Gamma(\beta /2 +1)\Gamma(a+b +(n+j-1)\beta /2+2)} }
\end{array}\right. .$$

\proclaim{Lemma}
For future use{\rm ,} the following expressions
 \bean b_n^{(\beta=1)}&:=&\frac{(n!)^2}{(n-2)!(n+2)!}
\frac{I^{(1)}_{n-2}I^{(1)}_{n+2}}{(I_n^{(1)})^2}
=\left\{
\begin{array}{l}
 \frac{n(n-1)}{16b^2}  ~~~~\mbox{ \rm(Gauss)}\\  \\
 \frac{n(n-1)(n+2a)(n+2a+1)}{16b^4}\\ \hspace{2cm}
\mbox{ \rm(Laguerre)}\\
\navs
 \frac{Q}{Q_6^{\pm}} ~~~~~~~~~~~  \mbox{\rm
(Jacobi)}\\
\end{array}\right.\\  && \\
\navs
 b_n^{(\beta=4)}&:=&\frac{(n!)^2}{(n-1)!(n+1)!}
\frac{I^{(4)}_{n-1}I^{(4)}_{n+1}}{(I_n^{(4)})^2}
  =\left\{
\begin{array}{l}
 \frac{2n(2n+1)}{4b^2}  ~~~~\mbox{ \rm(Gauss)}\\  \\
 \frac{2n(2n+1)(2n+a)(2n+a-1)}{b^4}\\ \hspace{2cm}
\mbox{ \rm(Laguerre)}\\
\navs
 \frac{Q}{Q_6^{\pm}} ~~~~~~~~~~~  \mbox{
{\rm (Jacobi)}}\\
\end{array}\right.
 \eean
 satisfy the following functional dependence\/{\rm :}\/
 $$
b_n^{(4)}(n,a,b)=b_{n}^{(1)}
 \left(-2n,-\frac{a}{2},-\frac{b}{2}\right).
$$
In the expressions above{\rm ,} $Q$ {\rm (}\/already
appearing in {\rm (0.3.1)),}
 and a new expression $Q_6^{\pm}$ are
expressible in terms of the variables $q,r,s$
introduced in {\rm (0.3.1):}
  \bean
 Q&:=& \left\{  \begin{array}{ll} 48(n-1)n(2a+n)(2a+n+1
 )(2b+n)\left(2b+n+1\right) \\~~~~ \left(2b+2a%
 +n+1\right)\left(2b+2a+n+2\right),&\mbox{for} 
 ~ (\beta=1) \\ \\
   96n\left(2n+1\right)\left(a+2\,n-1\right)\,\left(a+2n%
 \right)\,\left(b+2n-1\right)\\ ~~~~\left(b+2\,n\right)\,\left(b+a+2\,n-%
 2\right)\,\left(b+a+2\,n-1\right),&\mbox{for}~ 
  (\beta=4)\end{array}\right. \\
  &=&\frac{3}{16}\left(
(s^{2}-qr+q^{2})^{2}-4(rs^{2}-4q
 s^{2}-4s^{2}+q^{2}r)\right)
\\
  \eean
and\/\footnote{$\sqrt{q+1}=2n+2b+2a+1$ for $\beta=1$ and
 $\sqrt{q+1}=4n+b+a-1$ for $\beta=4$}

\smallbreak
  \bean Q^{\pm}_6
 &\hskip-5pt =\hskip-5pt&\left\{\begin{array}{ll}
\!\! = 48\,\left(b+a+n\right)\,\left(b+a+n+1\right)^{2}\left(b+a+n+2%
 \right)(2b+2a+2\,n-1)\\  ~~~~ 
 (2b+2a+2n+1
 )^{2}\,\left(2b+2a+2\,n+3\right),&\hskip-.75in  \hbox{for }\beta =1 \\ \navs
\!\! =3(b+a+4n-4)(b+a+4n-3)(b+a+4
 n-2)^{2}\\ ~~~~(b+a+4n-1)^{2}  (b+a+4n)
 (b+a+4n+1),& \hskip-.75in \hbox{for }\beta =4  \\ \end{array}
 \right.\\
 &\hskip-5pt=\hskip-5pt& 3
 q\left(q+1\right)\left(q-3\right)\left(q+4\pm4\sqrt{q+1}
 \right)~~~~~~~~  \left\{\begin{array}{lll} +& \mbox{for}& \beta=1\\
                             - & \mbox{for} & \beta=4
                             \end{array}\right.\ .\\
 \eean
 \endproclaim

 \demo{Proof}
 For instance, in the Jacobi case, one computes
\bean \frac{I^{(1)}_{n+2}}{I^{(1)}_{n}}&=& 2^{2n+2a
+2b+3}\ \ 
 \frac{\Gamma(\frac{n+3}{2}+a+b)
 \Gamma(\frac{n+4}{2}+a+b)
 \Gamma(\frac{n+2}{2}+a)
 \Gamma(\frac{n+2}{2}+b)
} {\Gamma(n+a+b+\frac{3}{2} )
 \Gamma(n+a+b+2 )
}  \\
 &&\cdot\ \frac{ \Gamma(\frac{n+3}{2}+a)
 \Gamma(\frac{n+3}{2}+b)
  \Gamma(\frac{n+3}{2})
   \Gamma(\frac{n+4}{2})}
   { \Gamma(n+a+b+\frac{5}{2} )
 \Gamma(n+a+b+3 )}\\
 \frac{I^{(4)}_{n+1}}{I^{(4)}_{n}}&=& 2^{4n+a +b}\ \ 
 \frac{\Gamma(2n+a+b)\Gamma(2n+a+1)
 \Gamma(2n+b+1)\Gamma(2n+3)}
 {\Gamma(4n+a+b )\Gamma(4n+a+b +2)}
 \eean
and so,
 \bean
\left.\frac{(n!)^2}{(n-2)!(n+2)!}
\frac{I^{(\beta)}_{n-2}I^{(\beta)}_{n+2}}
{(I^{(\beta)}_n)^2}\right|_{\beta=1}&=&
 \left.\frac{Q}{Q_6^{\pm}}\right|_{\beta=1}\\ \navs
 \left.\frac{(n!)^2}{(n-1)!(n+1)!}
\frac{I^{(\beta)}_{n-1}I^{(\beta)}_{n+1}}
 {({I^{(\beta)}_n})^2}\right|_{\beta=4}&=&
\left.\frac{Q}{Q_6^{\pm}}\right|_{\beta=4}.\\
\noalign{\vskip-36pt}
 \eean
\enddemo

\section{Proof of Theorems 0.1, 0.2, 0.3}

From Theorems 2.1 and 2.2, the integrals $I_n(t,c)$
, depending on  $\beta=2,1,4$, on $t=(t_1,t_2,...)$ and on
 the boundary points
$c=(c_1,...,c_{2r})$ of $E$, relate to
$\tau$-functions, as follows:
 \bea
 I_n(t,c)& =&\int_{E^n}|\Dt_n(z)|^{\beta}\prod_{k=1}^n
\left(e^{\sum_1^{\iy}t_i z_k^i}\rho(z_k)dz_k\right)
\spn{4.0.1}\\  \navs &=&\left\{
\begin{array}{lll}
 n!\tau_n(t,c),& n~\mbox{arbitrary}, &\beta=2
 \\
n!\tau_n(t,c),& n~\mbox{even}, &\beta=1
 \\ n!
\tau_{2n}(t/2,c), & n~\mbox{arbitrary},&\beta=4.
\end{array}
\right.\nn \eea
  $I_n(t)$ refers
to the integral (4.0.1) over the full range. It also follows that
$\tau_n(t,c)$ satisfies the KP-like equation\footnote{Remember $\delta^\beta_{1,4}=1$ for $\beta=1,4$, and $=0$ for
$\beta=2$.}
 \be
 12\frac{\tau_{n-2}(t,c)\tau_{n+2}(t,c)}{\tau_n(t,c)^2}
\delta^{\beta}_{1,4}=({\rm KP})_t\log \tau_n(t,c),~~
\left\{\begin{array}{l}
n~\mbox{arbitrary for} ~\beta=2\\
 n~\mbox{even for}~ \beta=1,4
   \end{array}\right. \spn{4.0.2}
    \ee
 where
  $$
{\rm (KP)}_t  F:= \left(\left(\frac{\pl}{\pl t_1}
\right)^4+3\left(\frac{\pl}{\pl
t_2}\right)^2-4\frac{\pl^2}{\pl t_1 \pl
t_3}\right)F+6\left(\frac{\pl^2}{\pl
t^2_1}F \right)^2.
 $$
\phantom{veron}

  \demo{{\rm 4.1.} $\beta=2,1$}
   Evaluating
the left-hand side of (4.0.2)(for $\beta=1$) yields,
taking into account $P_n:=P_n(E)=I_n(0,c)/I_n(0)$:
 \bean
\left.12\frac{\tau_{n-2}(t,c)\tau_{n+2}(t,c)}{\tau_n(t,c)^2}
\right|_{t=0}&=&12\frac{(n!)^2}{(n-2)!(n+2)!}
\left.\frac{I_{n-2}(t,c)I_{n+2}(t,c)}{I_n(t,c)^2}
\right|_{t=0}\\ \navs &=&12\frac{n(n-1)}{(n+1)(n+2)}
\frac{I_{n-2}(0)I_{n+2}(0)}{I_{n}(0)^2}
 \frac{P_{n-2}P_{n+2}}{P_n^2}
  \\ \navs & =&12b_n^{(1)}\frac{P_{n-2}(E)P_{n+2}(E)}{P_n^2(E)}
,\eean
 with $b^{(1)}_n$ given by Lemma 3.1. Concerning the
 right-hand side of (4.0.2), it follows from Section
 2.1 that
 $F_n(t;c)=\log I_n(t;c)$, as in (4.0.1), satisfies
Virasoro constraints, corresponding precisely to the
situation of Sections 3.1 and 3.2 for Gauss, Laguerre
 and Jacobi. As explained in (3.1.4), (3.2.1) and (3.2.2),
 we express
  $$
 \left.\frac{\pl ^4 F}{\pl t_1^4}\right|_{t=0},
 \left.\frac{\pl^2 F}{\pl t_2^2}\right|_{t=0},
 \left.\frac{\pl^2 F}{\pl t_1\pl t_3}\right|_{t=0},
  \left.\frac{\pl^2 F}{\pl t_1^2}\right|_{t=0},
~~~~~F=\log I_n(t,c),$$ in terms of ${\cal D}_k$ and
${\cal B}_k$, which when substituted in the right-hand side
of (4.0.2), i.e., in the KP-expressions, leads to
(upon comparing the expressions (1.3.4) and (1.3.8)
with (3.1.1) for Gauss and Laguerre and using (3.2.1)
directly for Jacobi): \bigbreak

\noindent $\bullet$ {\rm Gauss}$
\mbox{ with}\left\{\begin{array}{l}
\ga_{1,-1}=-\frac{1}{2},\ga_{1,0}=\ga_1=0,\dt_1=-\frac{n}{2}\\  \navs
\ga_{2,-1}=0,\ga_{2,0}=-1/2,\ga_{2,1}=0,
\ga_2=-\frac{n}{4} \sigma_1 
 ,\dt_2=0\\  \navs
\ga_{3,-1}=-\frac {1}{4} \sigma_1
,\ga_{3,0}=0,\ga_{3,1}=-\frac{1}{2},\ga_{3,2}=\ga_3=0,
\dt_3=-\frac{n}{4}\sigma_1
 .
\end{array}\right.
$
 \bean
  \lefteqn{ \left.{\rm (KP)}_t \log
  \tau_n(t,c)\right|_{t=0}}\\  \navs
&=&({\cal D}_1^4+6n{\cal D}_1^2+3{\cal D}_2^2-3{\cal
D}_2-4{\cal D}_1{\cal D}_3)F+6({\cal D}^2_1
F)^2+\frac{3}{4}(2-\beta)n(n-1)\\  \navs
 &=&
\frac{1}{16} \Bigl( ({\cal B} _{-1}^4+8(n+(2-\beta
)(n-1)){\cal B}_{-1}^2+12{\cal B}_0^2+24 {\cal
B}_0-16{\cal B}_{-1}{\cal B}_1)F
\\ &&\hspace{4cm}+6({\cal B}^2_{-1} F)^2
+12(2-\beta)n(n-1)\Bigr) \eean
\bigbreak 
\noindent $\bullet$  Laguerre 
  with $\left\{\begin{array}{l}
\ga_{1,-1}=0,\ga_{1,0}=-1,\ga_1=-\frac{n}{2}(\sigma_1+a),\\ \navs
\ga_{2,-1}=0,\ga_{2,0}=-\sigma_1,\ga_{2,1}=-1,
 \ga_2=-\frac{n}{2}\sigma_1(\sigma_1+a)
,\\ \navs \ga_{3,-1}=0,\ga_{3,0}=-\sigma_1 \sigma_2
,\ga_{3,1}=-\sigma_2,\\ \navs \ga_{3,2}=-1, \ga_3=
 -\frac{n}{2}\sigma_1\sigma_2(\sigma_1+a) .
\end{array}\right.
$

\begin{eqnarray*}
 \lefteqn{ \left.{\rm (KP)}_t \log
  \tau_n(t,c)\right|_{t=0}}\\ \navs
&=&\Biggl({\cal D} _1^4-2(\beta-3){\cal D}_1^3    \\ \navs
  &&~~ -\left(2n(n-1)(\beta-2)(\beta-1)+(\beta-2)(4an+4n+5)
  -4n^{2}-4an-1\right)
{\cal D}_1^2 \\ \navs &&
 ~~
-3\left(\beta-2\right)\left(\beta n^{2}-\beta n+2an+2n+1%
 \right) {\cal D}_1\\ \navs &&~~+3{\cal
D}^2_2 +6(\beta(n-1)+a+2) {\cal D}_2-6{\cal
D}_3-4{\cal D}_1 {\cal D} _3\Biggr)F_n\\ \navs
&&-3(\beta-2)({\cal D}_1F_n)^2+6({\cal
D}_1^2\log\tau_N)^2- 4(\beta-3)({\cal D}_1F_n)({\cal
D}^2_1F_n)\\ \navs
&& -\frac{3}{4}(\beta-2)n(n-1)(\beta n-2\beta+2a+%
 2)(\beta n-\beta+2a+2)
\end{eqnarray*}
\begin{eqnarray*}
&=&\Bigl( {\cal B}_{-1}^4 + 2(\beta -3){\cal B}_{-1}^3
 \\ \navs
&&- \left(\left(\beta-2\right)\left(3
 \left(\beta -1 \right)\left(n-1\right)^{2}
  +3n^{2}+6an-4\,a+2\right)+\left(a^2-1\right)\right) {\cal B}_{-1}^2
 \\ \navs
 &&+ 3\left(\beta-2\right)\left(
  \left(\beta-1\right)\left(n -1\right)^{2}+n^{2}+2an-a\right)
  {\cal B}_{-1}-4{\cal B}_{1}{\cal B}_{-1}-2{\cal B}_{1}
  \\ \navs
  &&+2\left(\beta n+a\right){\cal B}_{0}{\cal B}_{-1}+3
 {\cal B}^2_{0}-\left(\beta n+a\right){\cal B}_{0} \Bigr) F
\\ \navs
  &&
  +6({\cal B}_{-1}^{2}F)^2+4\left(\beta-
 3\right)({\cal B}_{-1}F)({\cal B}_{-1}^2F)+3\left(2-\beta\right)
 ({\cal B}_{-1}F)^{2} \\ \navs
  &&
 -\frac{3}{4}(\beta-2)n(n-1)(\beta n-2\beta+2a+%
 2)(\beta n-\beta+2a+2).
 \end{eqnarray*}

\bigbreak


 \noindent $\bullet$   Jacobi\footnote{In the
Jacobi $\beta=2$ case, we have $b_0=a-b,~b_1=a+b$; 
thus $
 r=2(b_0^2+b_1^2), q_n= 2(2 n+a +b )^2
$
and $q(q^2-4)=16(2n+\gamma+\delta)^2
 (2n+\gamma+\delta-1)(2n+\gamma+\delta +1)$.}

\bigbreak

  for $\beta=2$,

\medbreak

 \noindent $\frac{1}{8}q(q^2-4)\left.{\rm (KP)}_t
\log
  \tau_n(t,c)\right|_{t=0}$ \bean &=&\Bigl(2
{\cal B}_{-1}^4+(q-r+4){\cal B}^2_{-1}-(4{\cal
B}_{-1}F-s) {\cal B}_{-1}+3q {\cal B} _{0}^2 - 2q
{\cal B}_{0}+8{\cal B}_0 {\cal B}_{-1}^2\\
&&-\ 4(q-1){\cal B}_{1}{\cal B}_{-1} +(4{\cal
B}_{-1}F-s) {\cal B}_1 +2(4{\cal B}_{-1}F-s) {\cal
B}_{0} {\cal B}_{-1}+
 2 q {\cal B}_2 \Bigr)F
\\ && \hspace{2cm}
  +4{\cal B}^2_{-1}F\left( 2{\cal B}_{0}F
   +3{\cal B}^2_{-1}F\right)
\eean

for $\beta=1$, 

\medbreak

\noindent $ Q^{\pm}_6 \left.{\rm (KP)}_t \log
  \tau_n(t,c)\right|_{t=0}$
 \bean &=&(q+1)\Bigl(4q
{\cal
 B}^4_{-1}
 +12(4{\cal B}_{-1}F-s)
 {\cal B}^3_{-1}+2\left(q+12\right)(4{\cal B}_{-1}F-s)
 {\cal B}_{0}{\cal B}_{-1} \\
 &&+\ 3q^{2}
 {\cal B}_{0}^2 -4\left(q-4\right)q {\cal B}_{1}{\cal
B}_{-1}+q(4{\cal B}_{-1}F-s)  {\cal B}_{1}+20q{\cal
B}_{0}{\cal B}^2_{-1}+2q^{2}{\cal B}_{2}\Bigr) F
 \\
 &&+\ \Bigl(Q_2{\cal B}^2_{-1}-sQ_1{\cal
 B}_{-1}+Q_3{{\cal B}_0}\Bigr)F +48({\cal B}_{-1}F)^{4}\\
&&-\ 48s({\cal
B}_{-1}F)^{3}+2Q_4({\cal B}_{-1}F)^{2} \\
 &&+\ 12\,q^{2}({{\cal
B}_0}F)^{2}+16\,q\,\left(2\,q-1\right)
 {\cal B}^2_{-1}F{{\cal B}_0}F+24\left(q-1\right)q({\cal B}^2_{-1}F)^{2}
 \\
 &&+\ 24  \Bigl(2{\cal B}_{-1}F
 -s\Bigr){\cal B}_{-1}F\Bigl(  (q+2)
  {{\cal B}_0}F
 + (q+3){\cal B}^2_{-1}F\Bigr) +Q,
   \eean
where the $Q_1,Q_2,Q_3,Q_4,Q$ are given by (0.3.1) and
where the auxiliary $Q_6^{\pm}$ happens to be exactly
the one of Lemma 3.1. This establishes Theorems 0.1,
0.2 and 0.3 for $\beta=2,1$, at least when $b=1$ in
the exponent of the Gaussian and Laguerre ensembles,
upon noting that ${\cal B}^j_{k}\log P_n(E)=
 {\cal B}^j_{k}\log I_n(0,c)/I_n(0)=
 {\cal B}^j_{k}\log \tau_n(0,c)$.

Finally, a simple argument captures the case $b\neq
1$. Indeed, setting $\al
E:=\bigcup_1^{2r}~[\al~c_{2i-1},
 \al~c_{2i}]\subset
F $, for $\al>0$, the elementary identities
\begin{eqnarray*} I_n(t,c)&=&\int_{E^n}|\Dt_n(z)|^{\beta}\prod_{k=1}^n
e^{- bz_k^2}dz_k
 =C
 \int_{(\sqrt{b}~E)^n}|\Dt_n(z)|^{\beta}\prod_{k=1}^n
e^{- z_k^2}dz_k \\
I_n(t,c)&=& \int_{E^n}|\Dt_n(z)|^{\beta}\prod_{k=1}^n
z_k^{a} e^{-bz_k}dz_k
 = C
 \int_{({b}~E)^n}|\Dt_n(z)|^{\beta}\prod_{k=1}^n
z_k^{a} e^{- z_k}dz_k,
 \end{eqnarray*}
where $C(a,b,n,\beta)$ is a constant independent of $E$,
lead to the same Virasoro constraints as in Examples 1
and 2 (\S 1.3), but with the following mapping
for the differential
 operators
\bea
 \left({\cal B}_{-1},{\cal B}_{0},{\cal
B}_{1}\right)&\rightarrow&
 \left(\frac{{\cal B}_{-1}}{\sqrt{b}},{\cal B}_{0},{\cal
B}_{1}\sqrt{b}\right) ~~~\mbox{(Gauss)} \spn{4.1.1}\\
 &\rightarrow&
 \left({\cal B}_{-1},b{\cal B}_{0},b^2{\cal
B}_{1} \right)\, ~~~~\mbox{(Laguerre)}. \spn{4.1.2}
 \eea

Therefore, the equations (0.1.2) and (0.2.2) for the
probabilities (0.1.1) and (0.2.1) are obtained by
making the substitutions (4.1.1) and (4.1.2) in the
PDEs (0.1.2)$|_{b=1}$ and (0.2.2)$|_{b=1}$; this
process yields the precise equations (0.1.2) and
(0.2.2), with $b\neq 1$. This ends the proof of
Theorems 0.1, 0.2 and 0.3 for the cases $\beta=1,2$. \enddemo

 \demo{{\rm 4.2.} $\beta=4$, using duality}
From (4.0.1), the integral for $\beta=4$ is
expressible in terms of a $\tau$-function, in which
$t$ is replaced by $t/2$. Hence (4.0.2) becomes:
 \be
12\frac{\tau_{2n-2}(t/2,c)\tau_{2n+2}(t/2,c)}
 {\tau_{2n}(t/2,c)^2}
={\rm (KP)}_{t/2}(\log \tau_{2n})(t/2,c).\hskip.5in\spn{4.2.1}\ee
 So, the left-hand side of (4.2.1) equals
 ($P_n:=P_n(E)=I_n(0,c)/I_n(0)$)
  \bean
\left.12\frac{\tau_{2n-2}(t/2,c)
 \tau_{2n+2}(t/2,c)}{\tau_{2n}(t/2,c)^2}
\right|_{t=0}&=&12\frac{(n!)^2}{(n-1)!(n+1)!}
\left.\frac{I_{n-1}(t,c)I_{n+1}(t,c)}{I_n(t,c)^2}
\right|_{t=0}\\ \navs &=&
12 \frac{n }{ (n+1)}
\frac{I_{n-1}(0)I_{n+1}(0)}{I_n(0)^2}\frac{P_{n-1}P_{n+1}}{P_n^2}\\ \navs
 &=& 12b_n^{(4)}\frac{P_{n-1}(E)P_{n+1}(E)}{P_n^2(E)}
,\eean
 where $b_n^{(4)}=b_n^{(4)}(n,a,b)$ is given by
 Lemma 3.1 and satisfies
 $$
b_n^{(4)}(n,a,b)=b_{n}^{(1)}
 \left(-2n,-\frac{a}{2},-\frac{b}{2}\right).
$$
 Recall from Theorem 1.1  (1.1.4) that
$I_n^{(\beta)}(t,c;a_i,b_i)$ and $
I_n^{(4/\beta)}(t,c;a_i, b_i)$ (where we indicate the
explicit dependence on the coefficients $a_i$ and
$b_i$ of $\rho'/\rho$) satisfy the same equations,
with altered parameters:

$$\left({\cal B}_k-{\cal
V}^{(\beta)}_{k,n}(t;n,a_i,b_i)\right)
I_n^{(\beta)}(t,c;a_i,b_i)=0,$$

$$\left({\cal B}_k-{\cal
V}^{(\beta)}_{k,n}(-\frac{\beta
}{2}t;-\frac{2}{\beta}n, a_i,-\frac{\beta}{2}
b_i)\right) I_n^{(4/\beta)}(t,c;a_i, b_i)=0.$$

Setting $\beta=1$ in the equations above, extracting
$t$-partials in terms of ${\cal B}_k$'s, and using the
procedure explained in this section, we have that
\bean\left. {\rm (KP)}_{t}(\log
I_n^{(1)}(t,c;a_i,b_i))\right|_{t=0}&=&R({\cal
B};n,a_i,b_i) \log I_n^{(1)}(0,c;a_i,b_i)\\
 &=&R({\cal
B};n,a_i,b_i) \log P_n^{(1)}(E),
 \eean
\bean \left. {\rm (KP)}_{t/2}(\log I_n^{(4)}(t,c;a_i,b_i))
  \right|_{t=0}
& =&\left. {\rm (KP)}_{-t/2}(\log
I_n^{(4)}(t,c;a_i,b_i))\right|_{t=0}\\ &=& R({\cal
B};-2n,a_i,-b_i/2) \log I_n^{(4)}(0,c;a_i,b_i)\\
 &=&R({\cal
B};-2n,a_i,-b_i/2) \log P_n^{(4)}(E),
 \eean
where $R({\cal B};a_i,b_i,n)$ denotes the right-hand
side of the equations (0.1.2), (0.2.2) and (0.3.4) for $\beta=1$.
The coefficients $a_i$ and $b_i$ of the rational
function $-\rho'/\rho$ are as follows: the $a_i$ and
$b_i$ all vanish, except for

$$\begin{array}{llllll} {\rm Hermite}&a_0=1&a_1=0&a_2=0 &
b_0=0 & b_1=2b\\
{\rm Laguerre}&a_0=0&a_1=1&a_2=0&b_0=-a&b_1=b\\ {\rm Jacobi}&
a_0=1&a_1=0&a_2=-1&b_0=a-b&b_1=a+b;
\end{array}
 $$

\noindent thus the map $$
(n,a_i,b_i)\longrightarrow (-2n,a_i,-b_i/2)$$
translates into the map
 \be
  (n,a,b)\longrightarrow
(-2n,-a/2,-b/2), \spn{4.2.2}
 \ee
  which shows that the PDEs (0.1.2), (0.2.2) and (0.3.4)
   for the
case $\beta=4$ are obtained by means of the map
(4.2.2) from the same PDEs for $\beta=1$. But
according to (0.0.5), this is the precise way the
coefficients $q,s,Q_{-1},Q_0,Q_1,Q_2,\break Q_3,Q_4,Q$, evaluated at
$\beta =4$, are obtained from the same coefficients at
$\beta=1$. This ends the proof of Theorem 0.3.\hfill \qed
\enddemo

\demo{{\rm 4.3.} Reduction to Chazy and Painlev{\rm \'{\it e}} equations
$(\beta=2)$}
Setting $E=[-\iy,x], E=[0,x], E=[-1,x]$ in the PDEs
(0.1.2), (0.2.2) and (0.3.4) respectively, leads to
the equations (0.4.1), (0.4.2) and (0.4.3)
respectively, as announced in Section 0.4. Furthermore
setting $\beta=2$, the inductive terms on the left-hand side of
(0.4.1) and  (0.4.2)  vanish  and one obtains
the ODEs:

\begin{itemize}
  \item {\rm Gauss}: $\displaystyle{P_n(\max_i \lb_i \leq
  x)=\exp ({-\int_x^{\iy} f(u) du}})$, where $f$
  satisfies:
$$
 f^{\prime\prime\prime} +  6
~f^{\prime 2}+4b (2n-bx^2) f^{\prime} + 4b^2 x~f =0.$$
\item[]

  \item
  {\rm Laguerre}: $\displaystyle{P_n(\max_i \lb_i \leq
  x)=\exp \left(-\int_x^{\iy} \frac{f(u)}{u} du
  \right) }$, where $f$
satisfies
\begin{eqnarray*}
\hspace{-1cm} x^2f^{\prime\prime\prime} +
xf^{\prime\prime}
 + 6
xf^{ \prime 2} - 4 ff^{\prime}-((a-bx)^2 - 4nbx )
f^{\prime}- b(2n + a - bx) f = 0 .
\end{eqnarray*}
\item[]
  \item {\rm Jacobi}: $\displaystyle{P_n(\max_i \lb_i \leq
  x)=\exp \left(-\int_x^{1} \frac{f(u)}{1-u^2} du
  \right) }$, where $f$
 satisfies:

  \bean &&\hspace{-2cm}{ 2(x^2-1)^2f^{\prime\prime\prime}
 +4(x^2-1)\left(xf^{\prime\prime}
 -3f^{\prime 2}\right)
 +\left(16
xf-q_n(x^{2}-1)-2sx-r \right)f^{\prime}  }
 \nonumber\\&&\hspace{5cm}-f\left(4f-q_nx-s \right)=0,
 \eean

 \hspace{-1.1cm} where $r,s,q_n$ are defined in (0.3.1).

\end{itemize}

 These three equations are of the form
 \be
f^{\prime \prime\prime}+\frac{P'}{P}f^{
\prime\prime}+\frac{6}{P}f^{\prime
2}-\frac{4P'}{P^2}ff' +\frac{P^{\prime\prime}}{P^2}
f^2
+\frac{4Q}{P^2}f'-\frac{2Q'}{P^2}f+\frac{2R}{P^2}=0, \hskip.25in\spn{4.3.1}
\ee
 with the following coefficients $P,Q,R$:
  $$
\begin{array}{llll}
{\rm Gauss} & P(x)=1&~4Q(x)=-4b^2x^2+8bn&~R=0 \\  \navs  {\rm Laguerre} &
P(x)=x&~4Q(x)=-(bx-a)^2+4bnx
&~R=0
\\ \navs  {\rm Jacobi} &
P(x)=1-x^2&~4Q(x)=-\frac{1}{2}(q_n(x^2-1)+2sx+r)&~
 R=0 .\end{array}
 $$

The general Chazy class of differential equations are
equations of the form
 $$
f^{\prime\prime\prime}=F(z,f,f^{\prime},
 f^{\prime\prime}),~\mbox{where
$F$ is rational in $f,f^{\prime},f^{\prime\prime}$ and
locally analytic in z,}
 $$
  subjected to the requirement
that the general solution be free of movable branch
points; the latter is a branch point whose location
depends on the integration constants. In his
classification, Chazy found thirteen cases, the first
of which is given by (4.3.1), with arbitrary
polynomials $P(z), Q(z), R(z)$ of degree $3,2,1$
respectively.

 Cosgrove
(\cite{C}, \cite{CS}), (A.3), shows this third-order equation
 has a first integral, which is second-order in $f$
and quadratic in $f^{\prime\prime}$, \bea &f^{
\prime\prime 2}& +\frac{4}{P^2}
 \left( (Pf^{\prime 2}+Q f^{\prime}+R)f^{\prime}
 - (P' f^{\prime 2}+\frac{}{}Q' f^{\prime}+R')f^{}
  \right.\spn{4.3.2} \\ \navs && \hspace{2cm} \left.
   +\frac{1}{2}(P^{\prime\prime}f^{\prime
}+Q^{\prime\prime} )f^2
 -\frac{1}{6} P^{\prime\prime\prime}f^3 +c\right)=0,  \nonumber
\eea
  with an integration constant $c$. In the three cases,
  discussed above, $c=0$. Notice equations of the general form
  $$ f^{ \prime\prime
2}=G(x,f,f^{ \prime})$$ are invariant under the map $$
x\mapsto \frac{a_1z+a_2}{a_3z+a_4}~~ \mbox{and}~~
f\mapsto \frac{a_5f+a_6z+a_7}{a_3z+a_4}.$$ Using this
map, the polynomial $P(z)$ can be normalized to $$
P(z)=z(z-1),~z,~\mbox{or} ~1.$$

 Equation (4.3.2) is a master
Painlev\'e equation, containing the six Painlev\'e
equations. If $f(x)$ satisfies  the first three equations above,
then the new function $g(z)$, defined below,

$$
\begin{array}{ll}
{\rm Gauss} & g(z)=b^{-1/2}f(zb^{-1/2})+\frac{2}{3}nz\\ \navs
{\rm Laguerre} &
g(z)=f(z)+\frac{b}{4}(2n+a)z+\frac{a^2}{4}\\ \navs {\rm Jacobi} &
g(z):=-\frac{1}{2}f(x)|_{x=2z-1}-\frac{q}{8}z+\frac{q+s}{16}\\
\navs
 \end{array}
 $$
satisfies the following canonical equations of Cosgrove  and Scoufis ([11], [12]):
 \begin{itemize}
  \item $  g^{\prime\prime 2}=-4g^{\prime 3
}+4(zg^{\prime}-g)^2+A_1g^{\prime}+A_2$,
\hspace{3cm}({\rm Painlev\'e IV})
 
  \item $ (zg^{\prime\prime })^2=(zg^{\prime}-g)\Bigl(-4g^{\prime 2
}+A_1(zg^{\prime}-g)+A_2\Bigr)+A_3g^{\prime}+A_4
 $,

 \hspace{10cm}({\rm
Painlev\'e V})
 
  \item $ (z(z-1)g^{\prime\prime
})^2=(zg^{\prime}-g)\Bigl(4g^{\prime
2}-4g^{\prime}(zg^{\prime}-g)
 +A_2\Bigr) +A_1g^{\prime
2}+A_3g^{\prime}+A_4 $

\hspace{10cm}({\rm Painlev\'e VI})
\end{itemize}
with respective coefficients

\begin{itemize}
  \item
  $A_1=3\left( \frac{4n}{3}\right)^2,~A_2=-\left(
  \frac{4n}{3}\right)^3, $

\item
$
 A_1=b^2,~A_2=b^2((n+\frac{a}{2})^2+\frac{a^2}{2})
 ,~A_3=-a^2b(n+\frac{a}{2}),~
 A_4=\frac{(ab)^2}{2}\\
 \phantom{hu}~~~~~~~~~~~~~~~~~~~~~~~~~~~~~~~~~~~~~~~~~ 
 ~~~~~~~~~~~~~~~~~
 ~~~~ .((n+\frac{a}{2})^2+\frac{a^2}{8}),
 \\
 $
  \item

$A_1=\frac{2q+r}{8},~A_2=\frac{qs}{16},~A_3=\frac{(q-s)^2
+2qr}{64},~ A_4=\frac{q}{512}(2s^2+qr).
$

\end{itemize}
Each of the equations above can be transformed into
the standard Painlev\'e equations.

\section{Appendix.  Self-similarity proof of\\ the
Virasoro constraints (Theorem 1.1)}

 Given the data (0.0.1) to (0.0.3), namely
$\rho=e^{-V}$ and $-\rho^{\prime}/\rho=V'=g/f
=\sum_0^{\iy}b_iz^i  / \sum_0^{\iy}a_iz^i$ and
$E=\bigcup_1^{r}~[c_{2i-1},c_{2i}]\subseteq F\subseteq
\BR,$ we show that the multiple integral
\be
I_n(t,c;\beta):=\int_{E^n}|\Dt_n(x)|^{\beta}\prod_{k=1}^n
\left(e^{\sum_1^{\iy}t_i x_k^i}\rho(x_k)dx_k\right),
~~ \mbox{for}~ n>0\hskip.5in \spn{5.0.1}
\ee 
satisfies the Virasoro
constraints of Theorem 1.1, using a (much less
conceptual!) self-similarity argument. Setting
 $$
 dI_n(x):=|\Dt_n(x)|^{\beta}\prod_{k=1}^n
\left(e^{\sum_1^{\iy}t_i x_k^i}\rho(x_k)dx_k\right),$$
we state the following lemma:

\proclaim{Lemma} The following variational
formula holds\/{\rm :}\/
 \be\left.\frac{d}{d\vr}dI_n (x_i\mapsto
x_i+\vr f(x_i)x_i^{k+1} )\right|_{\vr=0}
=\sum^{\iy}_{\ell=0}
 \left(a_{\ell}~{}^{\beta}\BJ^{(2)}_{k+\ell,n}
-b _{\ell}~{}^{\beta}\BJ^{(1)}_{k+\ell
+1,n}\right)dI_n. \spn{5.0.2}\ee
\endproclaim

\demo{Proof} Upon setting
 \begin{eqnarray}
 E(x,t)&:=&\prod_1^n e^{\sum_{i=1}^{\iy}t_i
x_k^i}\rho(x_k)\spn{5.0.3}
\\
&=&\prod_1^n e^{-V(x_k,t)}~,~\mbox{
where} ~V(x,t):=V(x)-\sum_1^{\iy} t_i x^i,\nn\end{eqnarray}
 the following two relations hold:
 \bea
&&\spn{5.0.4}\\
  \left(\frac{1}{2}\sum_{{i+j=k}\atop{i,j>0}}\frac{\pl^2}{\pl
t_i\pl t_j}-\frac{n}{2} \delta_{k,0}\right)E
&=&\left(\sum_{{1\leq a<\beta\leq n
}\atop{{i,j>0}\atop{i+j=k}}}x^i_{\al}x^j_{\beta}+
\frac{k-1}{2}\sum_{1\leq\al\leq n}x_{\al}^k\right)E,
\nn \\ \navs \left(\frac{\pl}{\pl
t_k}+n\delta_{k,0}\right)E&=&\left( \sum_{1\leq\al\leq
n}x_{\al}^k\right)E,~~\mbox{for all $k\geq 0$}. \nn\eea So,
the point now is to compute the $\vr$-derivative
\be
\frac{d}{d\vr}\left(|\Delta_n(x)|^{\beta}
 e^{\sum^n_{k=1}(-V(x_k)+\sum^{\iy}
_{i=1}t_ix^i_k)} dx_1...dx_n\right)_{x_i\mapsto
x_i+\vr f(x_i)x_i^{k+1}}\Biggl|_{\vr=0}, \spn{5.0.5}\ee which
consists of three contributions:

\medbreak

  {\it Contribution} 1:

\medbreak
\noindent (5.0.6)

\noindent$\displaystyle{\frac{\pl}{\pl\vr}\left|\Delta(x+\vr
f(x)x^{k+1})\right|^{\beta}\Biggl|_{\vr=0}}$ \bea
&=&\beta|\Delta(x)|^{\beta}\sum_{1\leq\al <\ga\leq
n}\frac{\pl}{\pl\vr}\log\left(|x_{\al}-x_{\ga}+\vr(f(x_{\al})x_{\al}^{k+1}
-f(x_{\ga})x_{\ga}^{k+1})|\right)\Biggl|_{\vr=0}\nn \\ \navs 
&=&\beta|\Delta(x)|^{\beta}\sum_{1\leq\al <\ga\leq
n}\frac{f(x_{\al})x_{\al}^{k+1}-f(x_{\ga})x_{\ga}^{k+1}}
{x_{\al}-x_{\ga}}\nonumber\\ \navs 
&=&\beta|\Delta(x)|^{\beta}\sum_{\ell=0}^{\iy}a_{\ell}
\sum_{1\leq\al <\ga\leq n}
\frac{x_{\al}^{k+\ell+1}-x_{\ga}^{k+\ell+1}}{x_{\al}-x_{\ga}}\nonumber\\ \navs 
&=&\beta|\Delta(x)|^{\beta}\sum_{\ell=0}^{\iy}a_{\ell}
 \left(\sum_{{i+j=\ell+k}\atop{{i,j>0}\atop {1\leq\al
<\ga\leq
n}}}x^i_{\al}x^j_{\ga}+(n-1)\sum_{1\leq\al\leq
n}x_{\al}^{\ell+k}-\frac{n(n-1)}{2}\delta_{\ell+k,0}\right)\nonumber\\ \navs 
&=&\beta
E^{-1}|\Delta(x)|^{\beta}\sum_{\ell=0}^{\iy}a_{\ell}\Biggl(\frac{1}{2}
\sum_{{i+j=k+\ell}\atop{i,j>0}}\frac{\pl^2}{\pl t_i\pl
t_j}-\frac{n}{2}\delta_{k+\ell,0}\nonumber\\ \navs  &
&+\left(n-\frac{k+\ell+1}{2}\right)
\left(\frac{\pl}{\pl
t_{k+\ell}}+n\delta_{k+\ell,0}\right)-\frac{n(n-1)}{2}
\delta_{k+\ell,0}\Biggr)E\nonumber\\ \navs 
 &=&\beta
E^{-1}|\Delta(x)|^{\beta}\sum_{\ell=0}^{\iy}a_{\ell}
 \nonumber\\ \navs &&
 \Biggl(\frac{1}{2}
\sum_{{i+j=k+\ell}\atop{i,j>0}}\frac{\pl^2}{\pl t_i\pl
t_j}+\left(n-\frac{k+\ell+1}{2}\right) \frac{\pl}{\pl
t_{k+\ell}} +\frac{n(n-1)}{2}
\delta_{k+\ell,0}\Biggr)E.\nonumber   \eea

 \bigbreak
 {\it Contribution} 2:
\begin{eqnarray}
&&{\frac{\pl}{\pl\vr}\prod^n_1d(x_{\al}+\vr
f(x_{\al})x_{\al}^{k+1})\Biggl|_{\vr=0}}  \spn{5.0.7}\\ \navs
&&\hskip.25in =\sum^n_1\left(f'(x_{\al})x_{\al}^{k+1}
+(k+1)f(x_{\al})x_{\al}^k\right)\prod^n_1dx_i\nonumber\\
&&\hskip.25in=\sum^{\iy}_{\ell=0}(\ell+k+1)a_{\ell}\sum^n_{\al=1}x_{\al}^{k+\ell}
\prod^n_1dx_i\nonumber\\
&&\hskip.25in=E^{-1}\sum^{\iy}_{\ell=0}(\ell+k+1)a_{\ell}\left(\frac{\pl}{\pl
t_{k+\ell}}+n\delta_{k+\ell,0}\right)E \prod^n_1
~dx_i,\nonumber  \end{eqnarray}

\bigbreak
{\it Contribution} 3:
 \bea
&&\hskip-.5in{\frac{\pl}{\pl\vr}\prod^n_{\al=1}\exp
\left(-V \left(x_{\al}+\vr
f(x_{\al})x^{k+1}\right)\phantom{\sum^y}\right.}\spn{5.0.8}\\ \navs
&&\left. \phantom{\exp\Big(}+\sum^{\iy}_{i=1}t_i
\sum^n_{\al=1}\left( x_{\al}+\vr f(x_{\al})
x_{\al}^{k+1}\right)^i\right)\Biggl|_{\vr=0}\nn\\ \navs 
&=&\left(-\sum^n_{\al=1}V'(x_{\al})f(x_{\al})x_{\al}^{k+1}+\sum^{\iy}_{i=1}
it_i\sum^n_{\al=1}f(x_{\al})x_{\al}^{i+k}\right)
E\nonumber\\ \navs 
&=&\left(-\sum^{\iy}_{\ell=0}b_{\ell}\sum^n_{\al=1}
x_{\al}^{k+\ell+1}+\sum_{{\ell\geq 0}\atop{i\geq
1}}a_{\ell}it_i\sum_{\al=1}^nx_{\al}^{i+k+\ell}\right)
E\nonumber\\ \navs 
&=&\Biggl(-\sum^{\iy}_{\ell=0}b_{\ell}\left(\frac{\pl}{\pl
t_{k+\ell+1}}+n\delta_{k+\ell+1,0}\right)
 \nonumber\\ \navs  &&\quad
 +\sum^{\iy}_{\ell=0}a_{\ell}
\sum^{\iy}_{i=1}it_i\left(\frac{\pl}{\pl
t_{i+k+\ell}}+n\delta_{i+k+\ell,0}\right)\Biggr) E.
\nonumber \eea As mentioned, to conclude (5.0.2), we
must add up the three contributions (5.0.6), (5.0.7)
and (5.0.8), resulting in:
 \bea
&&\spn{5.0.9}\\
 \lefteqn{\left.\frac{\pl}{\pl\vr}dI_n
 (x_i\mapsto
x_i+\vr f(x_i)x_i^{k+1} )
 \right|_{\vr=0}}\nn \\ \navs 
&&\hskip-14pt = \left(\sum^{\iy}_{\ell=0}a_{\ell}\left(\frac{\beta}{2}J^{(2)}_{k+\ell}
+(n\beta +(\ell+k+1)(1-\frac{\beta}{2}))
J^{(1)}_{k+\ell}\right.\right.\nonumber\\ \navs  &
&\quad\quad\quad\quad
\left.\left.+n(( n-1)\frac{\beta}{2}
+1)\delta_{k+\ell,0}\right)-\sum^{\iy}_{\ell=0}b_{\ell}\left(
J^{(1)}_{k+\ell+1}+n\delta_{k+\ell+1,0}\right)\right)
dI_n(x).\nonumber \eea 
where   $J^{(i)}_{k}
 :=~{}^{\beta}J^{(i)}_{k}$, as in (1.1.8).  Thus we use (1.1.8) to end  the
proof of Lemma 5.1.\enddemo

\demo{Proof of Theorem {\rm 1.1}}  The change of
integration variable
$
x_i\mapsto x_i+\vr f(x_i)x_i^{k+1}
  $
   in the
integral (5.0.1) leaves the integral invariant, but it
induces a change of limits of integration, given by
the inverse of the map above; namely the $c_i$'s in
$E={\bigcup^r_1}[c_{2i-1},c_{2i}]$, get mapped as
follows: $$ c_i\mapsto c_i-\vr
f(c_i)c_i^{k+1}+O(\vr^2). $$ Therefore, setting
$$E^{\vr}=\displaystyle{\bigcup^r_1}[c_{2i-1}-\vr
f(c_{2i-1}) c_{2i-1}^{k+1}+O(\vr^2),c_{2i}-\vr
f(c_{2i})c_{2i}^{k+1} +O(\vr^2)],$$ we find, using
Lemma 5.1 and the fundamental theorem of calculus,
\begin{eqnarray*}
0&=&\frac{\pl}{\pl\vr}\int_{(E^{\vr})^{2n}}|\Delta_{2n}(x+\vr
f(x)x^{k+1})|\prod^{2n}_{i=1}e^{-V(x_i+\vr
f(x_i)x_i^{k+1} ,t)}d(x_i+\vr f(x_i)x_i^{k+1})\\ \navs 
&=&\left(-\sum^{2r}_{i=1}c_i^{k+1}f(c_i)\frac{\pl}{\pl
c_i} +\sum^{\iy}_{\ell=0}\left( a_{\ell}
~{}^{\beta}\BJ^{(2)}_{k+\ell,n}-b_{\ell}~{}^{\beta}\BJ^{(1)}_{k+\ell+1,n}
\right)\right)  I_{n}(t,c,\beta).
\end{eqnarray*}
This ends the alternative proof of Theorem 1.1.\enddemo

\bye

\bibitem{AvM1} M. Adler and P. van Moerbeke,
Matrix integrals, Toda symmetries, Virasoro
constraints and orthogonal polynomials, {\it Duke Math. J\/}.\
{\bf 80} (1995), 863--911.


\bibitem{AvM2} M. Adler and P. van Moerbeke,
The spectrum of coupled random matrices, {\it Ann.\ 
of Math\/}.\ {\bf 149} (1999), 921--976.

\bibitem{AvM4}  M. Adler and P. van Moerbeke,
 Integrals over classical groups, random permutations,
 Toda and Toeplitz lattices,
  {\it Comm. Pure Appl. Math\/}., to appear;
  http://front.math.ucdavis.edu/math.CO9912143.

 \bibitem{AvM5} M. Adler and P. van Moerbeke,
 The Pfaff lattice, matrix integrals and a map
from Toda to Pfaff, {\it Duke Math J\/}.\ (2000), to appear.
 (solv-int/9912008)

  \bibitem{AvM6} M. Adler and P. van Moerbeke,
Generalized orthogonal polynomials, discrete KP
 and Riemann-Hilbert problems, {\it Comm.\ Math.\ Phys\/}.\
  {\bf 207} (1999),  589--620.

 \bibitem{AHV} M. Adler, E. Horozov, and P. van Moerbeke,
  The Pfaff lattice and skew-orthogonal polynomials,
 {\it Internat.\  Math.\  Res.\  Not\/}.\ {\bf 11} (1999),
 569--0588.

 \bibitem{ASV1} M. Adler, T. Shiota, and P. van Moerbeke,
 Random matrices, vertex operators and the
Virasoro algebra, {\it Phys. Lett\/}.\ {\bf A 208} (1995), 67--78;

\bibitem{ASV2} M. Adler, T. Shiota, and P. van Moerbeke,
 Random matrices, Virasoro algebras and
noncommutative KP, {\it Duke Math.\ J\/}.\ {\bf 94} (1998), 379--431.

\bibitem{ASV3} M. Adler, T, Shiota, and P. van Moerbeke,
 Pfaff $\tau$-functions, {\it Math.\ Ann\/}.\   (2000), to appear.
(solv-int/9909010)

\bibitem{A} K. Aomoto, Jacobi polynomials associated with
Selberg integrals, {\it SIAM J.\ Math.\ Anal\/}.\  {\bf  18}
(1987), 547--549.



\bibitem{Awata} H. Awata, Y. Matsuo, S. Odake, and J. Shiraishi,
  Collective field theory, Calogero-Sutherland
model and generalized matrix models, {\it Phys.\ Lett.\ B} {\bf 347}
(1995), 49--55.

\bibitem{C} C. M. Cosgrove, Chazy classes IX-XII of third-order
differential equations, {\it Stud. Appl. Math\/}.\ {\bf 104}
  (2000), 171--228.

\bibitem{CS} C. M. Cosgrove and  G. Scoufis,  Painlev\'e classification of a
class of differential equations of the second order
and second degree, {\it Stud.\  Appl.\ Math\/}.\ {\bf 88}
 (1993), 25--87.

\bibitem{ Dickey} L. Dickey, {\it Soliton Equations and Hamiltonian 
Systems\/}, World Scientific Publ.\ Co., River Edge, NJ  (1991).

\bibitem{Haine} L. Haine and  J.\ P. Semengue,  The Jacobi
 polynomial ensemble and the Painlev\'e VI equation,
 {\it J.\  Math.\ Phys\/}.\ {\bf 40} (1999),  2117--2134.

\bibitem{Mahoux} G. Mahoux and   M.\ L. Mehta,   A method
of integration over matrix variables: IV, {\it  J.\ Physique\/} I 
{\bf 1} (1991), 1093--1108.

\bibitem{MehtaMahoux}   M.\ L. Mehta and  G. Mahoux,  Level
spacing functions and nonlinear differential
equations, {\it  J.\ Physique\/} I  {\bf 3} (1993), 697--715.

\bibitem{Mehta} M.\  L.\  Mehta, {\it Random Matrices\/}, 2nd ed.\
Academic Press, Boston, MA, 1991.

\bibitem{Mehta2} M.\ L. Mehta,   A nonlinear differential equation
and a Fredholm determinant, {\it J.\ Physique}  I,
(1992), 1721--1729.

\bibitem{Moore} G.\ Moore,  Matrix models of 2D gravity and
isomonodromic deformations, {\it Progr.\ Theor.\ Phys Suppl\/}.\
No.\ 102 (1990), 255--285.

\bibitem{TW1} C.\ A. Tracy and H.\ Widom, 
Fredholm determinants, differential equations and
matrix models, {\it Comm.\  Math.\ Phys\/}.\ {\bf 163} (1994),
33--72.

\bibitem{TW2} C.\ A. Tracy and H. Widom,  On orthogonal and
symplectic matrix ensembles, {\it Comm.\ Math.\ Phys\/}.\ {\bf
177} (1996), 727--754.

\bibitem{UT} K. Ueno and K. Takasaki,   Toda lattice
hierarchy, {\it Adv.\ Stud.\  Pure Math\/}.\ {\bf 4} (1984), 1--95.

\bibitem{vM1} P. van Moerbeke,  The spectrum of random matrices and
integrable systems, Group 21, in {\it Physical Applications
and Mathematical Aspects of Geometry, Groups and
Algebras\/}, Vol.\ II, 835--852 (H.-D. Doebner, W.
Scherer, and C. Schulte, eds.),  World Scientific Publ. Co., Singapore, 
1997.

 \bibitem{vM2} P. van Moerbeke,
 Integrable lattices:
  random matrices and random permutations,  Three lectures at the
  introductory workshop, Semester on ``Random matrices and
  statistical mechanics models", MSRI (1999).

\bibitem{Wig} E.\ P. Wigner,  On the statistical distribution
of the widths and spacings of nuclear resonance
levels, {\it Proc.\ Cambr.\ Philos.\  Soc\/}.\ {\bf 47} (1951),
790--798.

\end{thebibliography}

\bigskip

\centerline{(Received October 15, 1998)}
\centerline{(Revised  \ \ \ \ \ \ \ \ \ \ )}


\begin{references}

\bibitem{AvM1} 
\name{M. Adler} and \name{P. van Moerbeke},
Matrix integrals, Toda symmetries, Virasoro
constraints and orthogonal polynomials, {\it Duke Math. J\/}.\
{\bf 80} (1995), 863--911.

 
\bibitem{AvM2} \bibline,
The spectrum of coupled random matrices, {\it Ann.\ 
of Math\/}.\ {\bf 149} (1999), 921--976.



\bibitem{AvM4}  \bibline,
 Integrals over classical groups, random permutations,
 Toda and Toeplitz lattices,
  {\it Comm.\ Pure Appl. Math\/}. {\bf 54} (2001), 153--205 (math.CO9912143).

 \bibitem{AvM5} \bibline,
 The Pfaff lattice, matrix integrals and a map
from Toda to Pfaff, {\it Duke Math.\ J\/}.\ (2001), to appear 
 (solv-int/9912008).

  \bibitem{AvM6} \bibline,
Generalized orthogonal polynomials, discrete KP
 and Riemann-Hilbert problems, {\it Comm.\ Math.\ Phys\/}.\
  {\bf 207} (1999),  589--620.

 \bibitem{AHV} \name{M. Adler, E. Horozov}, and \name{P. van Moerbeke},
  The Pfaff lattice and skew-orthogonal polynomials,
 {\it Internat.\  Math.\  Res.\  Not\/}.\ {\bf 11} (1999),
 569--588.


 \bibitem{ASV1} \name{M. Adler, T. Shiota}, and \name{P. van Moerbeke},
 Random matrices, vertex operators and the
Virasoro algebra, {\it Phys.\ Lett\/}.\ A {\bf 208} (1995), 67--78.

 
\bibitem{ASV3} \bibline,
 Pfaff $\tau$-functions, {\it Math.\ Ann\/}.\   (2001), to appear 
(solv-int/9909010).





\bibitem{A} \name{K. Aomoto}, Jacobi polynomials associated with
Selberg integrals, {\it SIAM J.\ Math.\ Anal\/}.\  {\bf  18}
(1987), 545--549.


\bibitem{Awata} \name{H. Awata, Y. Matsuo, S. Odake}, and \name{J. Shiraishi},
  Collective field theory, Calogero-Sutherland
model and generalized matrix models, {\it Phys.\ Lett.\ } B {\bf 347}
(1995), 49--55.







\bibitem{C} \name{C. M. Cosgrove}, Chazy classes IX-XII of third-order
differential equations, {\it Stud.\ Appl.\ Math\/}.\ {\bf 104}
  (2000), 171--228.



\bibitem{CS} \name{C. M. Cosgrove} and  \name{G. Scoufis},  Painlev\'e classification of a
class of differential equations of the second order
and second degree, {\it Stud.\  Appl.\ Math\/}.\ {\bf 88}
 (1993), 25--87.



\bibitem{ Dickey} \name{L. Dickey}, {\it Soliton Equations and Hamiltonian 
Systems\/}, World Scientific Publ.\ Co., River Edge, NJ  (1991).






\bibitem{Haine} \name{L. Haine} and  \name{J.\ P. Semengue},  The Jacobi
 polynomial ensemble and the Painlev\'e VI equation,
 {\it J.\  Math.\ Phys\/}.\ {\bf 40} (1999),  2117--2134.



\bibitem{Mahoux} \name{G. Mahoux} and   \name{M.\ L. Mehta},   A method
of integration over matrix variables: IV, {\it  J.\ Physique\/} I 
{\bf 1} (1991), 1093--1108.



\bibitem{Mehta} \name{M.\  L.\  Mehta}, {\it Random Matrices\/}, 2nd ed.\
Academic Press, Boston, MA, 1991.

\bibitem{Mehta2} \bibline,   A nonlinear differential equation
and a Fredholm determinant, {\it J.\ Physique}  I,
(1992), 1721--1729.

\bibitem{MehtaMahoux}   \name{M.\ L. Mehta} and  \name{G. Mahoux},  Level
spacing functions and nonlinear differential
equations, {\it  J.\ Physique\/} I  {\bf 3} (1993), 697--715.


\bibitem{Moore} \name{G.\ Moore},  Matrix models of 2D gravity and
isomonodromic deformations, {\it Progr.\ Theor.\ Phys.\ Suppl\/}.,
No.\ 102 (1990), 255--285.



\bibitem{TW1} \name{C.\ A. Tracy} and \name{H.\ Widom}, 
Fredholm determinants, differential equations and
matrix models, {\it Comm.\  Math.\ Phys\/}.\ {\bf 163} (1994),
33--72.

\bibitem{TW2} \bibline,  On orthogonal and
symplectic matrix ensembles, {\it Comm.\ Math.\ Phys\/}.\ {\bf
177} (1996), 727--754.






\bibitem{UT} \name{K. Ueno} and \name{K. Takasaki},   Toda lattice
hierarchy, {\it Adv.\ Stud.\  Pure Math\/}.\ {\bf 4} (1984), 1--95.

\bibitem{vM1} \name{P. van Moerbeke},  The spectrum of random matrices and
integrable systems, Group 21, in {\it Physical Applications
and Mathematical Aspects of Geometry, Groups and
Algebras\/}, Vol.\ II, 835--852 (H.-D. Doebner, W.
Scherer, and C. Schulte, eds.),  World Scientific Publ. Co., Singapore, 
1997.

 \bibitem{vM2} \bibline,
 Integrable lattices:
  random matrices and random permutations,  in {\it Random Matrices and
  their Applications}, MSRI publications {\bf 40} (2001), 321--406.




\bibitem{Wig} \name{E.\ P. Wigner},  On the statistical distribution
of the widths and spacings of nuclear resonance
levels, {\it Proc.\ Cambr.\ Philos.\  Soc\/}.\ {\bf 47} (1951),
790--798.
\end{references}
\end{document}